\documentclass[final,3p,times,pdflatex]{elsarticle}
\usepackage{axodraw}
\usepackage{amsmath}
\usepackage{amssymb}
\usepackage{graphicx}
\usepackage{color} % LCT: for text editing
\usepackage{amsfonts} % LCT: for Germanic "e" and "m" to match \Re and \Im

\bibstyle{elsarticle-num}

\def\SOFTSUSY{{\tt SOFTSUSY}}
\def\NMSSMTools{{\tt NMSSMTools}}
\def\code#1{\small{\tt #1}\normalsize}

\newcommand{\be}{\begin{equation}}
\newcommand{\ee}{\end{equation}}
\newcommand{\ba}{\begin{eqnarray}}
\newcommand{\ea}{\end{eqnarray}}
\newcommand{\Zv}{\,\mathbf{\backslash}\mkern-11.0mu{\mathbb{Z}}_{3}} % AGW: Backslash looks better
\newcommand{\ds}{\displaystyle}
\newcommand{\overbar}[1]{\mkern 1.5mu\overline{\mkern-1.5mu#1\mkern-1.5mu}\mkern 1.5mu}
\newcommand{\lamsq}{\lambda^2}
\newcommand{\kapsq}{\kappa^2}
\newcommand{\tr}{\mathrm{Tr}}
\newcommand{\dt}{\frac{d}{dt}}
\newcommand{\mhusq}{m^2_{H_2}}
\newcommand{\mhdsq}{m^2_{H_1}}
\newcommand{\mlamsq}{M_\lambda^2}
\newcommand{\mkapsq}{M_\kappa^2}
\newcommand{\mssq}{m_S^2}
\newcommand{\mtrisq}{m_3^2}
\newcommand{\msprsq}{m_S'^2}
\newcommand{\mqsq}{m_{\tilde{Q}}^2}
\newcommand{\mdsq}{m_{\tilde{d}}^2}
\newcommand{\musq}{m_{\tilde{u}}^2}
\newcommand{\mlsq}{m_{\tilde{L}}^2}
\newcommand{\mesq}{m_{\tilde{e}}^2}
\newcommand{\Alam}{a_\lambda/\lambda}
\newcommand{\Akap}{a_\kappa/\kappa}
\newcommand{\Musq}{M_u^2}
\newcommand{\Mdsq}{M_d^2}
\newcommand{\Mesq}{M_e^2}

\def\at{\alpha_t}
\def\ab{\alpha_b}
\def\as{\alpha_s}
\def\atau{\alpha_{\tau}}
\def\oatab{O(\at\ab)}
\def\oatas{O(\at\as)}
\def\oabas{O(\ab\as)}

\def\oatq{O(\at^2)}
\def\oabq{O(\ab^2)}
\def\oatauq{O(\atau^2)}
\def\oabatau{O(\ab \atau)}

\journal{Computer Physics Communications}

\begin{document}

\begin{frontmatter}

%%% Adelaide preprint number
\begin{flushright}
ADP-13-33/T853
\end{flushright}

\title{Next-to-Minimal SOFTSUSY}

\author[damtp]{B.C.~Allanach}
\author[adelaide]{P.~Athron}
\author[adelaide,bern]{Lewis~C.~Tunstall\corref{cor1}}
\ead{tunstall@itp.unibe.ch}
\cortext[cor1]{Corresponding author}
\author[dresden]{A.~Voigt}
\author[adelaide]{A.G.~Williams}
\address[damtp]{DAMTP, CMS, University of Cambridge, Wilberforce road, Cambridge, CB3
  0WA, United Kingdom}
\address[adelaide]{ARC Centre of Excellence for Particle Physics at 
the Tera-scale, School of Chemistry and Physics, University of Adelaide, 
Adelaide SA 5005 Australia}
\address[bern]{Albert Einstein Center for Fundamental Physics, Institute for Theoretical Physics, University of Bern, Sidlerstrasse 5, CH-3012 Bern, Switzerland}
\address[dresden]{Institut f\"ur Kern- und Teilchenphysik,
TU Dresden, Zellescher Weg 19, 01069 Dresden, Germany}

\begin{abstract}
  We describe an extension to the
  \SOFTSUSY~program that provides for the calculation of the sparticle spectrum in the
  {\em Next-to-Minimal} Supersymmetric Standard Model (NMSSM), where a chiral
  superfield that is a singlet of the Standard Model gauge group is added to
  the Minimal Supersymmetric Standard Model (MSSM) fields. Often, a $\mathbb{Z}_3$
  symmetry is 
  imposed upon the model. \SOFTSUSY~can calculate the spectrum in this
  case as well as the case where general $\mathbb{Z}_3$ violating (denoted as $\Zv$) terms are
  added to 
  the soft supersymmetry breaking terms and the superpotential. 
  The user provides a theoretical boundary condition for the couplings and
  mass terms of the singlet.
  Radiative electroweak symmetry breaking data along with
  electroweak and CKM matrix data are used
  as weak-scale boundary conditions. 
  The renormalisation group equations are solved
  numerically between the weak scale and a high energy scale using a nested
  iterative algorithm. 
  This paper serves as a manual to the
  NMSSM mode of the program, detailing the approximations and
  conventions used. 
\end{abstract}

\end{frontmatter}

\section{Program Summary}
\noindent{\em Program title:} \SOFTSUSY{}\\
{\em Program obtainable   from:} {\tt http://softsusy.hepforge.org/}\\
{\em Distribution format:}\/ tar.gz\\
{\em Programming language:} {\tt C++}, {\tt fortran}\\
{\em Computer:}\/ Personal computer.\\
{\em Operating system:}\/ Tested on Linux 3.x\\
{\em Word size:}\/ 64 bits.\\
{\em External routines:}\/ None.\\
{\em Typical running time:}\/ A few tenths of a second per parameter point.\\
{\em Nature of problem:}\/ Calculating supersymmetric particle spectrum and
mixing parameters in the next-to-minimal supersymmetric standard
model. The solution to the renormalisation group equations must be consistent
with boundary conditions on supersymmetry breaking parameters, as
well as on the weak-scale boundary condition on gauge 
couplings, Yukawa couplings and the Higgs potential parameters.\\
{\em Solution method:}\/ Nested iterative algorithm and numerical minimisation
of the Higgs potential. \\
{\em Restrictions:} \SOFTSUSY~will provide a solution only in the
perturbative regime and it
assumes that all couplings of the model are real
(i.e.\ $CP-$conserving). If the parameter point under investigation is
non-physical for some reason (for example because the electroweak potential
does not have an acceptable minimum), \SOFTSUSY{} returns an error message.\\
{\em CPC Classification:} 11.1 and 11.6.\\
{\em Does the new version supersede the previous version?:} Yes.\\
{\em Reasons for the new version:} Major extension to include the
next-to-minimal supersymmetric standard model.\\
{\em Summary of revisions:} Added additional supersymmetric and supersymmetry
breaking parameters associated with the additional gauge singlet. Electroweak
symmetry breaking conditions are significantly changed in the next-to-minimal
mode, and some sparticle mixing changes. An interface to \NMSSMTools~has also
been included. Some of the object structure has also changed, and the command
line interface has been made more user friendly.

\newpage

\section{Introduction}

While TeV-scale supersymmetric particles have not yet been found\footnote{In some
  cases, lower   bounds of 1 
TeV or more have been placed upon gluinos and squarks by LHC experiments.}
at the LHC~\cite{Aad:2013wta,CMSspart}, searches for them continue along with
continuing strong theoretical interest in supersymmetric (SUSY) models. 
This is a
testament to the 
theoretical successes of weak-scale supersymmetry: chiefly the resolution of
the technical hierarchy problem, improvement of the apparent unification of Standard
Model (SM) gauge couplings and the provision of a potential 
dark matter candidate. 
In order to pursue SUSY phenomenology, a long calculational chain is
required~\cite{Allanach:2008zn}. Typically, this chain begins with the
calculation of the 
supersymmetric spectrum, including the couplings of the various sparticles and
Higgs bosons. Currently, in the Minimal Supersymmetric Standard Model (MSSM), there
are several spectrum generators: {\tt
  ISASUSY}~\cite{Baer:1993ae}, {\tt 
  SOFTSUSY}~\cite{Allanach:2001kg}, {\tt   SPheno}~\cite{Porod:2003um}, {\tt
  SUSEFLAV}~\cite{Chowdhury:2011zr} and {\tt SUSPECT}~\cite{Djouadi:2002ze}. 
Information from these spectrum generators is then passed to other programs
(for example those that calculate decays, that simulate collider events, or
that calculate the thermal relic density of dark matter) via data in the SUSY
Les Houches Accord format~\cite{Skands:2003cj}.

Recently a boson was discovered in the CMS and 
ATLAS experiments at over the 5$-\sigma$
level~\cite{Aad:2012tfa,Chatrchyan:2012ufa} with properties consistent
with the SM Higgs boson. Using 4.8 fb$^{-1}$ of 7 TeV data and 20.7
fb$^{-1}$ of 8 TeV data, ATLAS measures the mass to be
$m_h=125.5\pm0.2^{+0.5}_{-0.6}$ GeV by combining the $H \rightarrow \gamma
\gamma$ and $H \rightarrow ZZ$ decay channels~\cite{ATLAS-CONF-2013-014}.
In CMS, these channels give the combined constraint $m_h=125.3 \pm0.4 \pm0.5$
GeV in 5.1 fb$^{-1}$ of 7 TeV data and 5.3 fb$^{-1}$ of 8 TeV data. 
In the MSSM, one can often obtain a CP even Higgs that couples in a similar
way to the Standard Model Higgs boson. At tree-level, its mass is bounded by
$m_{h^0} < M_Z$, at odds with the LHC experiments' mass measurements. 
However,
the radiative corrections to the CP even Higgs mass can be sizeable,
particularly those from stops. The corrections are larger if the stops are
heavy, and if they are heavily mixed. Indeed, the MSSM has enough
flexibility~\cite{Djouadi:2013lra} such that the experimental values of
$m_{h^0}$ are achievable with TeV-scale stops and large mixing. On the other
hand, these relatively heavy stops reintroduce the little hierarchy problem,
requiring cancellation (at the level of one in several tens) 
between apparently unrelated parameters in the MSSM Higgs potential. 
Thus, we have the well known correlation~\cite{Barbieri:1998uv} between a
higher Higgs mass $m_{h^0}>106$ GeV and a higher level of apparently unnatural
cancellation. In several
well-studied simple models of supersymmetry breaking mediation, the
problem is much exacerbated~\cite{Arbey:2011ab}. 

In order to reduce the unnatural cancellations implied by the Higgs mass
measurement~\cite{Delgado:2010uj,Ellwanger:2011mu,King:2012tr,Perelstein:2012qg,Gherghetta:2012gb}, 
one can augment the MSSM by a gauge singlet chiral
superfield~\cite{BasteroGil:2000bw,Ellwanger:2009dp,Maniatis:2009re}. This model is referred to as the 
Next-to-Minimal 
Supersymmetric Standard Model (NMSSM)~\cite{NMSSM}. We shall distinguish
between a 
version where an extra symmetry is assumed (often a $\mathbb{Z}_3$ symmetry) 
and a version where it is not
($\Zv$)~\cite{Delgado:2010uj,Ell08,Ross:2011xv,Ross:2012nr}. 
In the MSSM, (based on a two Higgs doublet version of the SM with
softly broken $N=1$ global supersymmetry)
the tree-level bound upon the Higgs mass comes from the fact that 
the quartic Higgs couplings are related to the electroweak gauge couplings by
supersymmetry. The Higgs potential is modified by the addition of a gauge
singlet, and the resulting lightest CP even Higgs boson can receive additional
positive corrections to its mass at tree-level. In addition, the neutral Higgs
potential (now a function of three fields rather than two in the MSSM) is
heavily modified, with associated potential reductions in the unnatural
cancellations. Along with other factors this had lead to considerable interest in the NMSSM
in the recent literature and benchmarks points with a $125$ GeV Higgs have already been proposed \cite{King:2012is}.
It is therefore essential for the research community
to have access to a variety of reliable computational tools to calculate the relevant NMSSM observables. 

As mentioned above in the MSSM case, the initial step in a calculational
chain is 
typically spectrum and couplings calculation. Currently, there is one
out-of-the-box package {\tt NMSPEC}~\cite{Ellwanger:2006rn} which calculates
the spectrum of the 
NMSSM, matching weak-scale data with theoretical boundary conditions on
supersymmetry breaking and Higgs potential parameters. However, one can also 
marry {\tt SARAH}~\cite{Staub:2009bi,Staub:2010jh,Staub:2012pb,Staub:2013tta} with {\tt
  SPheno}~\cite{Porod:2003um} in order to be 
able to 
calculate the spectrum after setting up the model.\footnote{This has also been done in some non-NMSSM contexts --- for a recent example see \cite{Bharucha:2013ela}.} The NMSSM was included in
an extended version of the SUSY Les Houches Accord~\cite{Allanach:2008qq} so
that this 
information may be passed to programs performing other calculations. For
instance, 
{\tt
  NMHDECAY}~\cite{Ellwanger:2005dv} is 
capable of calculating the NMSSM Higgs decays, and
{\tt NMSDECAY}~\cite{Muhlleitner:2003vg,Das:2011dg} calculates sparticle
decays. \code{PYTHIA}~\cite{Sjostrand:2007gs} is then capable of simulating
particle collisions in the NMSSM and, in addition, \code{micrOMEGAs}~\cite{Belanger:2008sj}
can calculate the thermal dark matter relic density.

Having several public spectrum generators for the MSSM has proved fruitful for
the community. As well as comparisons and bug-finding, the various generators
have different levels of approximations and are able to calculate in different
generalisations of the MSSM. For example, some are easier (or harder) to use for certain
assumptions about supersymmetry breaking mediation. 
The advantages of having
several supported, publicly available spectrum generators naturally also
extends to the 
case of the increasingly popular NMSSM.
The extension of {\tt SOFTSUSY} to include the
NMSSM will hopefully aid the
accuracy and feasibility of a variety of NMSSM studies. 

In the present paper, we focus on the recent components that have been added to
{\tt SOFTSUSY} in order to include the effects of the gauge singlet
superfield. Up-to-date versions of this manual (along with other {\tt
  SOFTSUSY} manuals) 
will be released along with the
code in the {\tt doc/}~subdirectory. The other manuals in this subdirectory detail the standard
$R-$parity conserving MSSM~\cite{Allanach:2001kg}, 
the $R-$parity violating MSSM~\cite{Allanach:2009bv} and the loop-level
neutrino mass 
computation in the $R-$parity violating MSSM~\cite{Allanach:2011de}.
The remainder of the paper proceeds as follows: in section~\ref{sec:notation}, we
introduce the NMSSM supersymmetric parameters and the soft supersymmetry
breaking 
parameters using our conventions.  In section~\ref{sec:calculation}, we describe
the algorithm 
employed to calculate the spectrum of masses and couplings of NMSSM
particles, detailing our level of approximation for various parts of the
calculation. More technical information is relegated to the appendices. In
section~\ref{sec:run}, we explain how to run the program. The class structure,
along with the data contained within each class, is shown in
section~\ref{sec:objects}. Finally, in section~\ref{sec:RGEs}, we reproduce
the renormalisation group equations of the NMSSM to two-loops including the
full 3 by 3 flavour structure. 

\section{NMSSM Parameters \label{sec:notation}}

In this section, we introduce the NMSSM parameters
in the \SOFTSUSY~conventions. The translations to the variable
names used in the program code are shown explicitly in
section~\ref{sec:objects}.  

\subsection{Supersymmetric parameters \label{susypars}}
The chiral superfield particle content of the NMSSM has the 
following $SU(3)_c\times SU(2)_L\times U(1)_Y$ quantum numbers
\begin{align}
L&:(1,2,-\tfrac{1}{2})\,, & \bar{E}&:(1,1,1)\,, & 
Q&:(3,2,\tfrac{1}{6})\,,  & \bar{U}&:(\overline 3,1,-\tfrac{2}{3})\,, \notag \\
\bar{D}&:(\overline 3,1,\tfrac{1}{3})\,, & H_1&:(1,2,-\tfrac{1}{2})\,, & 
H_2&:(1,2,\tfrac{1}{2})\,, & S&:(1,1,0)\,.
\label{fields}
\end{align}
$S$ is the gauge singlet chiral superfield that is particular to the NMSSM. 
$L$, $Q$, $H_1$, and $H_2$ are the left-handed doublet lepton and quark 
superfields and the two Higgs doublets. $\bar{E}$, $\bar{U}$, and $\bar{D}$ are 
the lepton, up-type quark and down-type quark right-handed superfield singlets, 
respectively. Note that the lepton doublet superfields $L^a_i$ and the Higgs 
doublet superfield $H_1$ coupling to the down-type quarks have the same 
SM gauge quantum numbers. We denote an $SU(3)$ colour 
index of the fundamental representation by  $\{x,y,z\} \in \{1,2,3 \}$. The 
$SU(2)_L$ fundamental representation indices are denoted by 
$\{a,b,c\} \in \{1,2\}$ and the generation indices by $\{i,j,k\} \in \{1,2,3\}$.
 $\epsilon_{xyz}=\epsilon^{xyz}$ and  $\epsilon_{ab}=\epsilon^{ab}$ are totally
antisymmetric tensors, with $\epsilon_{123}=1$ and $\epsilon_{12}=1$, 
respectively.  Currently, only real couplings in the superpotential and Lagrangian are 
included. 

The full renormalisable, $R-$parity conserving superpotential is given by
\begin{align} 
 W_{\Zv}  &=  \epsilon_{ab} \left[ (Y_E)_{ij} L_i^b H_1^a \bar{E}_{j} 
+ (Y_D)_{ij} Q_i^{bx} H_1^a \bar{D}_{jx} 
+ (Y_U)_{ij} Q_i^{ax} H_2^b \bar{U}_{jx} 
+ (\lambda S + \mu)(H^a_2 H^b_1) \right]  + \xi_FS 
+ \frac{\mu^\prime}{2} S^{2} + \frac{\kappa}{3}S^{3} \\
&= W_\mathrm{MSSM}^{\mu =0} 
+ \epsilon_{ab}\left[ (\lambda S + \mu)(H^a_2 H^b_1) \right]  
+ \xi_FS + \frac{\mu^\prime}{2}S^{2} + \frac{\kappa}{3} S^{3} \,
\label{eq:WZ3V}
\end{align}
\noindent where $(Y_{U,D,E})_{ij}$ and $\lambda,\kappa$ are dimensionless Yukawa 
couplings, $\mu$ and $\mu'$ are supersymmetric mass terms, and $\xi_F$ encodes 
the effects of the supersymmetric tadpole term.  We use
the subscript $\Zv$~to reflect the fact that this superpotential 
contains terms which violate the $\mathbb{Z}_3$ symmetry that is commonly 
imposed on the NMSSM.  Imposing the $\mathbb{Z}_3$ symmetry restricts the 
superpotential to
\begin{align} 
 W_{\mathbb{Z}_3} &= \epsilon_{ab} \left[(Y_E)_{ij} L_i^b H_1^a \bar{E}_{j} 
+ (Y_D)_{ij} Q_i^{bx} H_1^a \bar{D}_{jx} + (Y_U)_{ij} Q_i^{ax} H_2^b \bar{U}_{jx}  
+ \lambda S(H^a_2 H^b_1) \right] + \frac{\kappa}{3}S^{3} \\
&= W_\mathrm{MSSM}^{\mu =0}  + \epsilon_{ab} \lambda S (H^a_2 H^b_1) 
+ \frac{\kappa}{3}S^{3}.
\label{eq:WZ3C}
\end{align}
\noindent The $\mathbb{Z}_3$-NMSSM superpotential Eq.~(\ref{eq:WZ3C}) contains no 
explicit mass parameter, thereby allowing a solution to the $\mu$-problem when 
the singlet field acquires a Vacuum Expectation Value (VEV) and generates an 
effective $\mu$ term of the right size. As such, it is sometimes referred to as 
the scale invariant NMSSM in the literature.  In this paper, we will always 
write $\mathbb{Z}_3$-NMSSM for the $\mathbb{Z}_3$ conserving case 
Eq.~(\ref{eq:WZ3C}) and $\Zv$-NMSSM for the general $\mathbb{Z}_3$ violating one 
Eq.~(\ref{eq:WZ3V}). 

For parameters common to both the MSSM and either the $\mathbb{Z}_3$-NMSSM or 
$\Zv$-NMSSM, a comparison of the \SOFTSUSY~conventions and the literature can be 
found in Table 1 of the MSSM \SOFTSUSY~manual \cite{Allanach:2001kg}.  Elsewhere, 
our conventions are those of the SUSY Les Houches Accord~\cite{Allanach:2008qq} 
and thus consistent with the review of Ellwanger, Hugonie and Teixeira (EHT) 
 \cite{Ellwanger:2009dp} and also Ref.~\cite{Degrassi:2009yq}. (Note however that 
our definitions of the neutral Higgs VEVs (section~\ref{sec:hpot}) differ by a 
factor of $\sqrt{2}$ compared to Refs.~\cite{Ellwanger:2009dp,Degrassi:2009yq}.)

\subsection{Next-to-minimal SUSY breaking parameters \label{sec:susybreak}}
The soft breaking scalar potential is given by

\be 
V_{\textrm{soft}} = V_3 + V_2\big|^{}_{m_3^2=0} + m_S^2|S|^2 
+ \epsilon_{ab} \lambda A_\lambda S H^a_2 H^b_1 
+ \frac{\kappa A_\kappa}{3} S^3 + V_{\Zv} \,,
\ee 
where all $\Zv$~terms are included in
\be 
V_{\Zv} =  \xi_S S + \frac{m_S^{\prime \, 2}}{2} S^2
  + \epsilon_{ab} m_3^2 H_2^a H_1^b + \textrm{h.c.} \,.
\label{eq:VZ3V}
\ee
Expressions for the trilinear scalar interaction potential $V_3$ and scalar 
bilinear SUSY breaking potential $V_2$  of the MSSM are given in Sect.\ 2.2 of 
the \SOFTSUSY~manual \cite{Allanach:2001kg} for the $R$-parity conserving MSSM.
 The notation $V_2\big|_{m_3^2=0}$ indicates that the $\Zv$~soft bilinear mass 
$m_3^2$ present in $V_2$ is set to zero to avoid double counting with the third term in Eq.~(\ref{eq:VZ3V}).

\subsection{Higgs potential and electroweak symmetry breaking}\label{sec:hpot}
At tree-level, the Higgs potential is given by
\begin{align}
V_\mathrm{Higgs} &=  V^H_F + V^H_D + V_{\rm soft}^H  \\
&=  V^{\mu=0}_\mathrm{MSSM} + V^{HN}_{F} + V_{\rm soft}^{HN}\,,
\end{align} 
where 
\begin{align}
V^{HN}_{F} &=   |\lambda S + \mu|^2 (|H_2|^2+|H_1|^2) + |\lambda H_2H_1
+\kappa S^2 + \mu^\prime S + \xi_S|^2  \,, \label{eq:HpotF} \\
 V_{\rm soft}^{HN}  &=   m_S^2|S|^2
+ \Bigg(\lambda A_{\lambda}SH_2H_1+\frac{\kappa}{3} A_{\kappa}S^3+ \ds\frac{m_S^{\prime \, 2}}{2} S^2 + \xi_S S + \textrm{h.c.} \Bigg)\,.
\label{eq:HpotS} 
\end{align}

\noindent The three neutral Higgs fields then pick up VEVs
\be 
        \langle H_1^0 \rangle = \ds\frac{1}{\sqrt{2}}{v_1 \choose 0}\,, 
\qquad  \langle H_2^0 \rangle = \ds\frac{1}{\sqrt{2}}{0 \choose v_2}\,, 
\qquad  \langle S \rangle =  \ds\frac{1}{\sqrt{2}}s\,, 
\label{eq:potmin} 
\ee
\noindent which are related to the soft masses via the minimization conditions
\begin{align}
m_{H_1}^2&= -\frac{M_Z^2}{2}\cos(2\beta) - \ds\frac{\lamsq}{2} v_2^2
 + (m_3^2)_\textrm{eff} \tan\beta 
- |\mu_\textrm{eff}|^2\,, \label{eq:mind}\\
m_{H_2}^2&= \frac{M_Z^2}{2}\cos(2\beta) - \ds\frac{\lamsq}{2}v_1^2 
+  \frac{(m_3^2)_\textrm{eff}}{\tan\beta} 
- |\mu_\textrm{eff}|^2 \,, \label{eq:minu} \\
m_S^2 &= -\kappa^2 s^2 - \ds\frac{\lamsq}{2} 
v^2 + \kappa\lambda v_2v_1
+ \lambda A_{\lambda} \frac{v_2v_1}{\sqrt{2}s}
-\kappa A_{\kappa}s  - m^{\prime \,2}_S - \mu^{\prime \,2} + 2 \kappa \xi_F  - 3 \kappa s \mu^\prime \,,
\label{eq:mins}
\end{align}
where $M_Z^2 = \tfrac{1}{4}\bar{g}^2(v_1^2+v_2^2)$ and $\overline{g} = (g_2^2+g^{\prime 2})^{1/2}$ 
for gauge couplings $g_2$ and $g^{\prime}=\sqrt{3/5}g_1$ of $SU(2)_L$ and (unnormalised) 
$U(1)$ interactions respectively.
We have $\tan \beta = v_2 /  v_1$ and for simplicity we have introduced
\be (m_3^2)_\textrm{eff} \equiv
 \ds\frac{ \lambda s}{\sqrt{2}} B_\textrm{eff} + \widehat{m}_3^2\,, \ee and
\be  \mu_\textrm{eff} \equiv
 \mu + \frac{\lambda s}{\sqrt{2}}\,, \;\;\;\; B_\textrm{eff}\equiv A_\lambda+\ds\frac{\kappa s}{\sqrt{2}}\,, \;\;\;\; \widehat{m}_3^2 \equiv m_3^2 + \lambda \Bigg(\ds\frac{\mu^\prime s}{\sqrt{2}} + \xi_F\Bigg)\,. \ee

\subsection{Tree-level masses \label{sec:tree}}
The chargino and sfermion masses are obtained by substituting 
$\mu\to\mu_\textrm{eff}$ into the MSSM expressions. The neutralino mass matrix 
is contained in the Lagrangian term  
$-\frac{1}{2}{\tilde\psi^0}{}^T{\cal M}_{\tilde\psi^0}\tilde\psi^0$ + h.c., where 
$\tilde\psi^0 =$ $(-i\tilde b,$ 
$-i\tilde w^3,$ $\tilde h_1,$ $\tilde h_2, \tilde{s})^T$ and
\begin{equation}
{\cal M}_{\tilde\psi^0} \ =\ \left(\begin{array}{ccccc} 
M_1 & 0 &-M_Zc_\beta s_W & M_Zs_\beta s_W & 0 \\
 0 & M_2 & M_Zc_\beta c_W & -M_Zs_\beta c_W & 0 \\ 
-M_Zc_\beta s_W & M_Zc_\beta c_W & 0 & -\mu & -\lambda v_2 \\
M_Zs_\beta s_W & -M_Zs_\beta c_W & -\mu & 0 & - \lambda v_1 \\
0 & 0 & 0 & 0 & 2 \kappa s + \mu^\prime
\end{array} \right)\,. \label{mchi0}
\end{equation} 
We use $s$ and $c$ for sine and cosine, so that
$s_\beta\equiv\sin\beta,\ c_{\beta}\equiv\cos\beta$ and $s_W (c_W)$ is
the sine (cosine) of the weak mixing angle.  
The 5 by 5 neutralino mixing matrix is an orthogonal matrix $O$ with real 
entries, such that $O^T {\cal M}_{\tilde\psi^0} O$ is diagonal. The neutralinos 
$\chi^0_i$ are defined such that their absolute masses increase with increasing 
$i$. Note that some of their mass values can be negative. 

The CP-even gauge eigenstates $(H^0)^T = (H_1^0,\, H_2^0, \, S)$ are rotated into
 mass eigenstates $(h^0)^T = (h_1, h_2, h_3)$ by a mixing matrix $R$,
\be 
h^0 = R H^0\,. 
\ee 
The mass matrix $M^2_{H^0}$ is obtained by expanding $H_{1,2}$ and $S$ about their 
VEVs (\ref{eq:potmin}) and identifying terms $-(H^0)^T M^2_{H^0} H^0$ in the Lagrangian.  
Typically, the resulting matrix elements $(M_{H^0}^2)_{ij}$ are simplified by using the 
tree-level electroweak symmetry breaking (EWSB) conditions (\ref{eq:mind}-\ref{eq:mins}) 
in order to eliminate the soft terms $m_{H_1}^2$, $m_{H_2}^2$ and $m_S^2$.  This is 
equivalent to defining  
\be
(M_{H^0}^2)_{ij} \equiv  \ds\frac{\partial^2 V}{\partial v_i \partial v_j} 
- \ds\frac{\delta_{ij}}{v_i}\ds\frac{\partial V}{\partial v_i} \qquad \mbox{with } v_3\equiv s\,,
\ee
and under this prescription we find    
\ba
 (M_{H^0}^2)_{11} & = & M_Z^2 c_\beta^2 
 + \Bigg(\ds\frac{\lambda s}{\sqrt{2}} B_\textrm{eff} +
 \widehat{m}_3^2\Bigg)\,\tan\beta\,,\\
 (M_{H^0}^2)_{12} & = & (4\lambda^2 - \overline{g}^2) \ds\frac{v_2 v_1 }{4}- 
 \ds\frac{\lambda s}{\sqrt{2}} B_\textrm{eff} - \widehat{m}_3^2\,, \\ 
 (M_{H^0}^2)_{13} & = & \lambda \Bigg[2 \mu_\textrm{eff}\,\ds\frac{ v_1}{\sqrt{2}} -
 (B_\textrm{eff} + \kappa s + \mu')\ds\frac{ v_2}{\sqrt{2}}\Bigg]\,,\\
 (M_{H^0}^2)_{22} & = & M_Z^2 s_\beta^2 + \Bigg(\ds\frac{\lambda s}{\sqrt{2}} B_\textrm{eff} +
\widehat{m}_3^2\Bigg)/\tan\beta\, \\
 (M_{H^0}^2)_{23} & = & \lambda \Bigg[2 \mu_\textrm{eff}\, \ds\frac{ v_2}{\sqrt{2}} -
(B_\textrm{eff} + \frac{\kappa s}{\sqrt{2}} + \mu')\ds\frac{ v_1}{\sqrt{2}}\Bigg]\,, \\
 (M_{H^0}^2)_{33} & = & \ds\frac{\lambda}{\sqrt{2}} (A_\lambda + \mu') \frac{v_2 v_1}{s}
+ \frac{\kappa s}{\sqrt{2}} (A_\kappa + 4\frac{\kappa s}{\sqrt{2}}+ 3 \mu') - \sqrt{2}(\xi_S + \xi_F \mu')/s\,.
\label{eq:MH0}
\ea

The three imaginary components of the neutral Higgs fields 
$(H^I)^T = (H^I_1, H_2^I, S^I)$ mix to give the two physical CP odd bosons 
$A_{1,2}$ and the Goldstone boson $G^0$.  A mixing matrix $P$ relates the two 
bases
\be 
a = P H^I\,, 
\ee
where $a^T = (G^0,A_1,A_2)$.  Here, $P$ matches the conventions of 
\cite{Degrassi:2009yq}, while deleting the first row from $P$ produces the 2 by
 3 mixing matrix for the physical CP-odd Higgs bosons in SLHA2 conventions \cite{Allanach:2008qq}. Following EHT \cite{Ellwanger:2009dp}, the entries of the 
3 by 3 mass matrix $ M^{\prime \, 2}_{P}$ in the $H^I$ basis read
\ba
( M^{\prime \, 2}_{P})_{11} & = & \Bigg(\ds\frac{\lambda s}{\sqrt{2}} B_\textrm{eff} +
\widehat{m}_3^2\Bigg)\,\tan\beta , \\
( M^{\prime \, 2}_{P})_{12} & = & \ds\frac{\lambda s}{\sqrt{2}} B_\textrm{eff} +
\widehat{m}_3^2, \\
( M^{\prime \, 2}_{P})_{13} & = & \lambda v_u (A_\lambda - 2\kappa s - \mu'), \\
( M^{\prime \, 2}_{P})_{22} & = & \Bigg(\ds\frac{\lambda s}{\sqrt{2}} B_\textrm{eff} +
\widehat{m}_3^2\Bigg)/\tan\beta ,  \\
( M^{\prime \, 2}_{P})_{23} & = & \lambda v_d (A_\lambda - 2\kappa s - \mu')\\
( M^{\prime \, 2}_{P})_{33} & = & \lambda (B_\textrm{eff}+3\kappa s +\mu')\ds\frac{v_u
v_d}{s} -3\kappa A_\kappa s  -2 m_{S}'^2 -\kappa \mu' s 
-\xi_F\left(4\kappa + \frac{\mu'}{s}\right) -\ds\frac{\xi_S}{s}.
\label{eq:MA0}
\ea
where tree-level EWSB has been imposed.

Note that --- as in the MSSM --- the mixing of the Goldstone boson $G^0$ depends
 only on $\tan\beta$. As shown in EHT \cite{Ellwanger:2009dp}, this can be seen 
by first performing a rotation by $\beta$, which converts $M^{\prime\, 2}_P$ to be 
block diagonal.  The resulting 2 by 2 submatrix may then be diagonalised. 
Therefore the CP-odd mixing can be stored as a single mixing angle.%
  \footnote{\SOFTSUSY~does this internally by storing $\theta_{A^0}$ in the 
    {\tt sPhysical} object (see Eq.~(\ref{nmssmsoftsusy})).  Note that the
    SLHA output 
    gives the 3 by 2 mixing matrix and thus matches SLHA2 conventions.}

Finally, the charged Higgs fields in the mass basis contain one massless 
charged Goldstone boson $G^{\pm}$ and a charged Higgs, $H^\pm$ with mass
\be 
m_{H^\pm}^2 = \left(\ds\frac{\lambda s}{\sqrt{2}} B_\textrm{eff} +
\widehat{m}_3^2\right)(\tan \beta + \cot \beta) + M_W^2 - \ds\frac{\lambda^2 v^2}{2}\,. 
\ee
  
\section{Calculation Algorithm \label{sec:calculation}}
\begin{figure}
\begin{center}
\begin{picture}(323,245)
\put(10,0){\makebox(280,10)[c]{\fbox{7.\ Calculate Higgs and
      sparticle pole masses at $M_{SUSY}$. Run to $M_Z$.}}}
\put(10,40){\makebox(280,10)[c]{\fbox{6.\ Run to $M_Z$.}}}
\put(150,76.5){\vector(0,-1){23}}
\put(10,80){\makebox(280,10)[c]{\fbox{5.\ Run to $M_X$. Apply soft breaking
and NMSSM SUSY boundary conditions.}}}
\put(150,116.5){\vector(0,-1){23}}
\put(10,120){\makebox(280,10)[c]{\fbox{4.\ EWSB with iterative solution for $\mu_\textrm{eff}$, outputs $\{s,\kappa,m_S\}$ in $\mathbb{Z}_3$-NMSSM and $\{\mu, m_3^2, \xi_S \}$ in $\Zv$-NMSSM.}}}
\put(150,156){\vector(0,-1){23}}
\put(10,160){\makebox(280,10)[c]{\fbox{3.\ Run to $M_{SUSY}$.}}}
\put(30,170){convergence}
\DashLine(110,165)(-70,165){5}
\DashLine(-70,165)(-70,5){5}
\DashLine(-70,5)(10,5){5}
\put(10,5){\vector(1,0){2}}
\put(150,197){\vector(0,-1){24}}
\put(150,239){\vector(0,-1){26}}
\put(60,245){\fbox{1.\ SUSY radiative corrections to
$g_i(M_Z)$.}}
\put(10,200){\makebox(280,10)[c]{\fbox{2.SUSY radiative corrections to
$h_{t,b,\tau}(M_Z)$.}}} 
\put(182,45){\line(1,0){190}}
\put(371,45){\line(0,1){200}}
\put(371,245){\vector(-1,0){143}}
\end{picture}
\end{center}
\caption{Iterative algorithm used to calculate the NMSSM spectrum. 
The initial step is the
uppermost one. $M_{SUSY}$ is the scale at which the EWSB
conditions 
are imposed, as discussed in the text. $M_X$ is the scale at which the high
energy SUSY breaking boundary conditions are imposed. Although Higgs and
sparticle masses are calculated at $M_{SUSY}$, the empirical values of 
electroweak boson and quark/lepton masses are imposed at $M_Z$. It is the
\SOFTSUSY~convention to evolve $\overbar{DR}$ couplings to $M_Z$ as the final step,
although in the SLHA2 output~\cite{Allanach:2008qq}, various couplings at
$M_{SUSY}$ are output. 
\label{fig:algorithm}}
\end{figure}
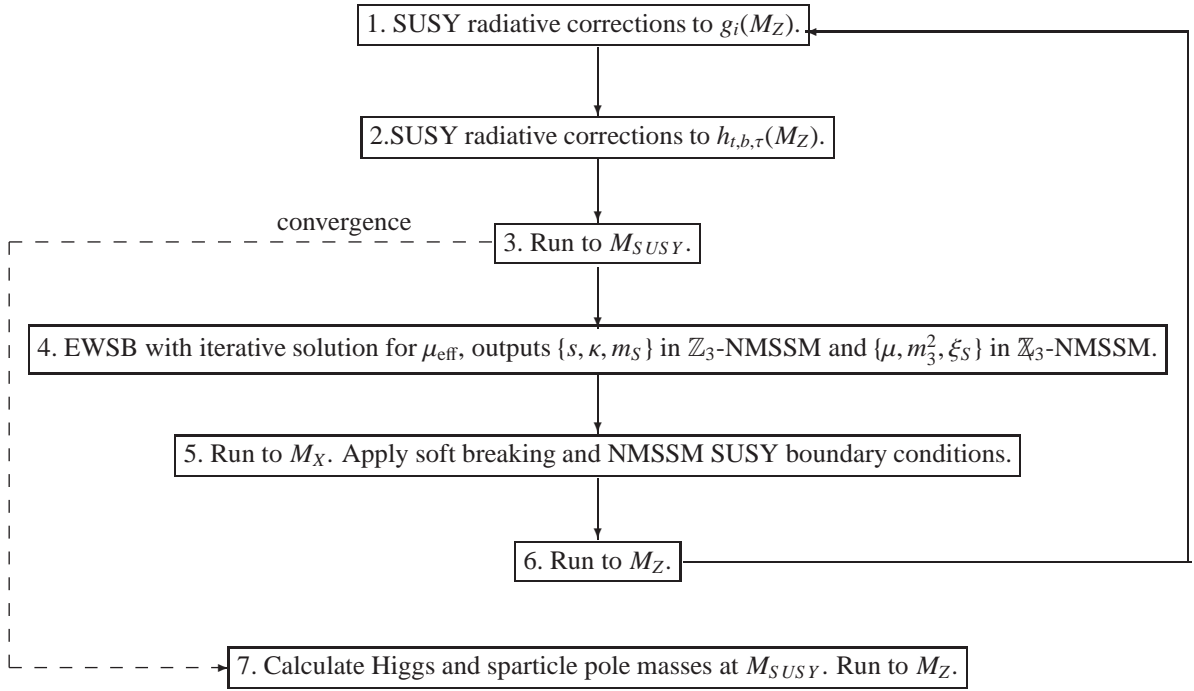

We now describe the algorithm used to perform the calculation.  The full 
iterative algorithm to determine the mass spectrum is shown schematically in 
Fig.~\ref{fig:algorithm}.  Here we will provide a detailed description of this 
procedure and specify all contributions that are included in the calculation.

As in MSSM \SOFTSUSY, the SM fermion and gauge boson masses, and the
 couplings $\alpha(M_Z)$, $G_F^\mu$, and $\alpha_s(M_z)$ act as low energy 
constraints. Below $M_Z$, the evolution of these input parameters proceeds in 
the manner described in Sect.\ 3.1 of the MSSM \SOFTSUSY~manual 
\cite{Allanach:2001kg}.  Similarly, the initial guess for the SUSY preserving 
$\overline{DR}$ parameters at $m_t$ follows the procedure outlined in Sect.\ 3.2
of \cite{Allanach:2001kg}, with the additional NMSSM parameters 
$\{\lambda, \kappa, s, \xi_F, \mu^\prime \}$ either initially set to their 
(user specified) input values, or to zero in the case when $\kappa$ and $s$ are 
treated as outputs from the EWSB conditions (section~\ref{ewsb}).

\subsection{Running of NMSSM couplings~\label{running}}
Following the initial guess at $m_t$, the two-loop $\beta$ functions of the 
$\Zv$-NMSSM are used to evolve the SUSY preserving parameters to a user 
specified scale $M_X$. If gauge unification has been specified as a boundary 
condition, $M_X$ is revised to leading-log order to provide a more accurate 
value upon the next iteration:
\begin{equation}
M_X^{\textrm{new}} = M_X \exp 
\left({\frac{g_2(M_X) - g_1(M_X)}{g_1'(M_X) - g_2'(M_X)}}\right)\,,
\label{mguteq}
\end{equation}
where primes denote derivatives calculated to two-loop order.   

In all stages of the calculation, the evolution of the NMSSM parameters is 
governed by three family, two-loop renormalization group equations (RGEs), whose
 form \cite{MV94,Yam94} for a general, $N=1$ semi-simple SUSY gauge 
theory is known. From these general results, it is possible to derive the 
explicit expressions of the RGEs in a chosen model (e.g.\ the work of Martin and
 Vaughn \cite{MV94} provides a complete list of the RGEs for the MSSM).  

In the case of the NMSSM considered here, it is a simple task to generalize the 
MSSM expressions \cite{MV94} to include contributions due to superpotential 
parameters such as $\lambda$ and their soft SUSY-breaking counterparts 
$a_\lambda$.  (Naturally, the RGEs for such parameters must be derived 
separately.)  The two-loop RGEs for the $§\Zv$-NMSSM are presented in the review by EHT \cite{Ellwanger:2009dp}, with  
the third family approximation
\begin{equation}
Y_U \approx \left(\begin{array}{c c c} 
0 & 0 & 0 \\
0 & 0 & 0 \\
0 & 0 & y_t 
\end{array}\right) \,, \qquad
Y_D \approx \left(\begin{array}{c c c} 
0 & 0 & 0 \\
0 & 0 & 0 \\
0 & 0 & y_b 
\end{array}\right) \,, \qquad
Y_E \approx \left(\begin{array}{c c c} 
0 & 0 & 0 \\
0 & 0 & 0 \\
0 & 0 & y_e 
\end{array}\right)\,,
\label{eq:3rd fam} 
\end{equation}
imposed to simplify the resulting expressions.
However, in \SOFTSUSY~the whole calculation is performed with quark 
flavor-mixing between all three families, so it is necessary to derive the 
additional NMSSM contributions from the general RGEs \cite{MV94,Yam94}.  The 
resulting expressions are collected in section~\ref{sec:RGEs} and in each case we have 
found agreement with the results of EHT \cite{Ellwanger:2009dp} once the third 
family approximation Eq.~(\ref{eq:3rd fam}) is enforced.  Note that in the 
\SOFTSUSY~conventions, all $\beta$ functions are real.
We also incorporate the two-loop running for $\tan\beta$ and the Higgs VEVs  
$v_{1,2}$ and $s$.  Here, we make use of the results obtained by 
Sperling et al.\ \cite{Sper13,Sper13-2}, where the pure   NMSSM contributions 
are reproduced in section~\ref{sec:RGEs}.
The program can be made to run faster by switching off the two-loop
 renormalization of the scalar masses and tri-linear scalar couplings.
Once the user-supplied boundary conditions are applied at $M_X$, the whole 
ensemble of NMSSM soft breaking and SUSY preserving couplings are evolved to 
$M_Z$. The inclusion of radiative corrections to the gauge and Yukawa couplings 
(steps 1 and 2 in Fig.~\ref{fig:algorithm}), and NMSSM renormalization (step 3)
 is analogous to MSSM \SOFTSUSY~--- for details we refer the reader to sections
3.3 and 3.4 of the \SOFTSUSY~manual \cite{Allanach:2001kg}.

\subsection{Low energy boundary conditions and electroweak 
symmetry breaking \label{ewsb}} 
The electroweak symmetry breaking (EWSB) conditions (\ref{eq:mind}-\ref{eq:mins}) 
allow one to constrain three model parameters of the theory.  With the central value 
for the $Z$ pole mass $M_Z$ taken as input, we rewrite Eqs.~(\ref{eq:mind}) and 
(\ref{eq:minu}) in terms of $\mu_\textrm{eff}^2$ and $(m_3^2)_\textrm{eff}$, as in 
the MSSM.  By including tadpole corrections $t_i$ and the transverse self energy $\Pi^T_{ZZ}$ of the 
$Z$ boson, we find
\begin{align}
  \mu_\textrm{eff}^2(M_{SUSY}) &=
  \frac{m_{\overline{H}_1}^2(M_{SUSY}) -
    m_{\overline{H}_2}^2(M_{SUSY}) \tan^2 \beta(M_{SUSY})}{\tan^2
    \beta(M_{SUSY}) - 1} - \frac{1}{2} M_{\overline Z}^2
  (M_{SUSY})\label{eq:mueffcond}\\ 
  (m^2_3)_\textrm{eff}(M_{SUSY})&=\frac{\sin{2\beta}(M_{SUSY})}{2}\Bigg\{\overline{m}_{H_u}^2(M_{SUSY})+\overline{m}_{H_d}^2(M_{SUSY})+
  2\mu_\textrm{eff}^2(M_{SUSY})\Bigg[1+\frac{\overline{M}_z^2}{\overline{g}^2s^2}(M_{SUSY})\Bigg]\Bigg\}\,,
  \label{eq:bmucond}
\end{align} 
where 
$m_{\overline{H}_i}^2 = m_{H_i}^2 - t_i/v_i$, 
$M_{\overline Z}^2(M_{SUSY}) = M_Z^2 + \Re\mathfrak{e}\Pi_{ZZ}^T(M_{SUSY})$ is the 
$\overline{DR}$ running (mass)$^2$ of the $Z$ boson.  Through Eqs.~(\ref{eq:mueffcond}) and (\ref{eq:bmucond}) we
can fix $\mu_\textrm{eff}$ and $(m^2_3)_\textrm{eff}$ in a similar manner
to the MSSM.  Note however, that in this case these are {\it effective} 
parameters constructed from several model parameters, so we must select which
of the latter are fixed.
In the $\mathbb{Z}_3$-NMSSM, we fix $s$ via Eq.~\ref{eq:mueffcond} and $\kappa$
via Eq.~\ref{eq:bmucond}, and use the third EWSB condition (\ref{eq:mins}) to fix
$m_S^2$.  In the $\Zv$-NMSSM, we have more freedom and can choose to fix
$\mu$ and $m_3^2$ --- as in the MSSM --- and use the third EWSB
condition to fix $\xi_S$.
Alternatively, the EWSB conditions (\ref{eq:mind}-\ref{eq:mins}) can be used to fix the soft Higgs
masses $m_{H_1}^2$, $m_{H_2}^2$ and $m_S^2$: see
\ref{sec:run}.

The full one-loop tadpole corrections
 from \cite{Degrassi:2009yq} are implemented, along with NMSSM two-loop 
$\oatas$ and $\oabas$ contributions \cite{Degrassi:2009yq} to the tadpoles.%
\footnote{We thank Pietro Slavich for kindly supplying us with the {\tt FORTRAN}
 files.}
 The two loop corrections from the MSSM are used for $\oatq$, $\oabatau$, 
$\oabq$, $\oatauq$ and $\oatab$, though it should be noted that these are not
complete in the NMSSM.  In both one-loop and two-loop cases, the tadpole 
corrections themselves depend on the output from the EWSB conditions, therefore 
an iteration is employed to find a self consistent solution.
After the EWSB iteration converges, the whole set of NMSSM parameters are run to
 $m_Z$. As detailed in Section 3.3 of \cite{Allanach:2001kg}, the gauge 
couplings $g_1$, $g_2$ and $g_3$ (where $g_1$ is the GUT normalised gauge 
coupling of $U(1)_Y$) and third family $\overline{DR}$ Yukawa couplings, $y_t$, 
$y_b$ and $y_\tau$ are fixed, including the precision corrections at $M_Z$.  Note
 however, that the expressions for the one-loop self energies of the gauge 
bosons and fermions are modified to match those given in \cite{Degrassi:2009yq}
 for the NMSSM.

\SOFTSUSY~calculates corrections to $\sin \theta_W$ following the procedure outlined 
in \cite{Pierce:1997zz}.  We use the same procedure in the NMSSM, with expressions 
for the MSSM self energies \cite{Pierce:1997zz} generalised to include NMSSM contributions 
\cite{Degrassi:2009yq}.  In the Higgs sector, we only consider contributions from the 
lightest NMSSM Higgs, since contributions from heavy Higgs states are negligible 
\cite{Pierce:1997zz}.  This is achieved by taking the Higgs state whose mass and coupling 
produces the contributions listed in \cite{Pierce:1997zz} once the MSSM limit is taken.  
Note that this ensures a simple MSSM limit for threshold corrections. 

In the $\Zv$-NMSSM, the parameters $\kappa$, $s$, $\xi_F$ and $\mu^\prime$ are 
reset to their input values at $M_Z$.  The parameters are then evolved back to 
$M_{\rm SUSY}$ where $M_Z^2$ and $\tan\beta$ are predicted as part of a consistency 
check.  If the user has specified that any of the parameters $\lambda$, $\kappa$, $s$, 
$\xi_F$ and $\mu'$  are to be input at the SUSY scale rather than the default option of 
inputting them at the GUT scale\footnote{See \ref{sec:run} for details on how to do this.} 
then they are set here.  

In general, the scalar Higgs potential (in both the $\mathbb{Z}_3$- and 
$\Zv$-NMSSM) can possess several local minima \cite{Ellwanger:2009dp}, so we 
include a test at $M_{\rm SUSY}$ to determine whether the chosen parameter space 
point corresponds to a global minimum (as done in the {\tt NMSPEC}~\cite{Ellwanger:2006rn} 
CHECKMIN routine).  The test works by comparing the value of the physical potential at 
the VEVs $v_u,v_d,s$ against scenarios where two or more VEVs are zero.  We 
include one-loop radiative corrections to the effective potential from third 
generation quarks and squarks; corrections from other sfermions are negligible 
due to their small Yukawa couplings. The parameters are then evolved back up to $M_X$ and 
the procedure is repeated until convergence is achieved, as shown in Fig.~\ref{fig:algorithm}. 
(If the iteration does not converge to the desired accuracy, \SOFTSUSY~outputs a 
{\tt No convergence} warning message --- see also Appendix C in \cite{Allanach:2001kg}.)

\subsection{NMSSM spectrum \label{spec}}
After the iteration has converged we calculate the pole masses.  The
Higgs pole masses are calculated using one-loop self energies from Degrassi and 
Slavich \cite{Degrassi:2009yq}, with additional $\Zv$~contributions to the 
triple Higgs couplings included (see Appendix A of EHT \cite{Ellwanger:2009dp}).
Two-loop corrections \cite{Degrassi:2009yq} of $\oatas$ and $\oabas$ are 
incorporated via {\tt FORTRAN} files provided by Pietro Slavich.  Contributions 
of order $\oatq$, $\oabatau$, $\oabq$, $\oatauq$ and $\oatab$ are included from 
the MSSM {\tt FORTRAN} files (also supplied by Pietro Slavich), but we note that
 these expressions receive additional NMSSM contributions which are currently 
unavailable.  Consequently, our calculation is not correct to this order, but 
rather to $\oatas$ and $\oabas$.  Nevertheless, the higher order MSSM 
contributions provide (a) a good approximation in the vicinity of the MSSM limit
, and (b) easier comparisons against MSSM results.

The sfermions, neutralinos and charginos also receive new NMSSM corrections to 
their self energies. To the best of our knowledge, the required expressions are 
presented only in \cite{Staub:2010ty}. However, we found a number of 
typographical errors in the published results \cite{Staub:2010ty}, whose 
origin%
\footnote{F.~Staub, private communication.} 
was due to the need to manually condense the auto-generated \LaTeX~output from 
{\tt SARAH} \cite{Staub:2009bi,Staub:2010jh,Staub:2012pb,Staub:2013tta}.  In 
particular, the self energy expressions generated by {\tt SARAH} do not contain 
these errors. Therefore, we used a combination of results listed in 
\cite{Staub:2010ty}, auto-generated \LaTeX~output from {\tt SARAH} for the self 
energies, plus individual checks of our own.
Finally, all one-loop self energies, tadpole corrections, and two-loop RGEs were
 unit tested against code pieces auto-generated from {\tt FlexibleSUSY} 
\cite{flexi-susy}, an in development {\tt MATHEMATICA} package for generating 
{\tt C++} code which makes use of the aforementioned {\tt SARAH} package.

\section*{Acknowledgments}
This work has been partially supported by STFC under grant number ST/J000434/1 
and by the Australian Research Council through its Centres of Excellence 
program. 

We thank Pietro Slavich for supplying us with the NMSSM {\tt FORTRAN} files with
two-loop $\oatas$ and $\oabas$ contributions to the Higgs masses and also a
{\tt FORTRAN} file with one-loop self energies (which we used as a cross check),
 as well as his helpful explanations on how to use them. We thank Florian 
Staub for responding quickly to our questions regarding \cite{Staub:2010ty}, and
 on questions and bug reports when comparing against {\tt SARAH} and 
{\tt FlexibleSUSY}. We also thank Ben Farmer for providing useful feedback after
 using a pre-release verion of the code.  

PA and AV thank Dominik St\"ockinger for many helpful comments and discussions 
regarding the precision corrections included here, and also thank both him and 
Jae-hyeon Park for listening to a number of discussions about this project in 
general and for offering helpful remarks. PA also thanks Roman Nevzorov for 
useful discussions.  LCT is supported by the Federal Commission for Scholarships
 for Foreign Students (FCS).  

\appendix

\section{Running \SOFTSUSY}
\label{sec:run}

\SOFTSUSY~produces an executable called \code{softpoint.x}. For the calculation
of the spectrum of single points in parameter space, we recommend the
SUSY Les Houches Accord 2 (SLHA2)~\cite{Allanach:2008qq}  input/output
option. The user must provide a file (e.g.\ the example file included
in the \SOFTSUSY~distribution
\code{rpvHouchesInput}), that specifies the model dependent input
parameters. The program may then be run with
\small
\begin{verbatim}
 ./softpoint.x leshouches < nmssmHouchesInput
\end{verbatim}
\normalsize

NMSSM-\SOFTSUSY\ accepts input files compliant with the SLHA2 format
given in Ref.~\cite{Allanach:2008qq} and supports the setting of all
SLHA2 input blocks associated with non-complex couplings.  The set of
input parameters which also exist in the MSSM are entered as described
in \cite{Skands:2003cj}, just as for the MSSM version of \SOFTSUSY,
while the new NMSSM parameters are all given in the \code{EXTPAR}
block as outlined in \cite{Allanach:2008qq}.  For example, in the
$\Zv$-NMSSM one can set:
\begin{verbatim}
Block EXTPAR                 # Z3 violating NMSSM
#  23   100                  # mu
#  24   1000                 # m_3^2 / (cos(beta) * sin(beta))
   61   0.1                  # lambda(MX)
   62   0.1                  # kappa(MX)
   63   1000                 # A_lambda(MX)
   64   1000                 # A_kappa(MX)
   65   500                  # (lambda * <S>)(MX)
   66   100                  # xi_F(MX)
#  67   1000                 # xi_S(MX)
   68   1000                 # mu'(MX)
   69   1000                 # m'_S^2(MX)
   70   1000                 # m_S^2(MX)
\end{verbatim}
The parameters
\begin{equation}
  \{\mu,m_3^2 / (\cos\beta \sin\beta),\xi_S\}: \qquad \Zv\mbox{-NMSSM} \label{eq:Z3v_ewsb}
\end{equation}
must not be set here, because they are output by the EWSB conditions, see
section \ref{ewsb}.  In the $\mathbb{Z}_3$-NMSSM one must set
all $\mathbb{Z}_3$ violating parameters ($\mu$, $m_3^2$, $\xi_F$,
$\xi_S$, $\mu'$, $m_S^{\prime\, 2}$) to zero or comment them
out.%
  \footnote{Unset parameters are assumed to be zero.}
The parameters
\begin{equation}
  \{\kappa,s,m_S^2\}: \qquad \mathbb{Z}_3\mbox{-NMSSM} \label{eq:Z3_ewsb}
\end{equation}
are then output from the EWSB conditions, as in section \ref{ewsb}, and they should
therefore not be set either.  One is then left with the following three
free parameters:
\begin{verbatim}
Block EXTPAR                 # Z3 symmetric NMSSM
   61   0.1                  # lambda(MX)
   63   1000                 # A_lambda(MX)
   64   1000                 # A_kappa(MX)
\end{verbatim}
By default all parameters are input at the scale $M_X$, defined either by (a) entry $0$ in
block \code{EXTPAR} or (b) as the gauge coupling unification scale where 
$g_1$ = $g_2$ (determined iteratively) when entry $0$ in block \code{EXTPAR} 
is not set.  

Should the user desire to input the parameters $\lambda$, $\kappa$, $\lambda s / \sqrt{2}$, 
$\xi_F$ and $\mu'$ at $M_{SUSY}$, a corresponding $-1$
entry in the block \code{QEXTPAR} has to be given:
\begin{verbatim}
Block QEXTPAR
   61   -1                   # input lambda at Msusy
   62   -1                   # input kappa at Msusy
   65   -1                   # input lambda * <S> at Msusy
   66   -1                   # input xi_F at Msusy
   68   -1                   # input mu' at Msusy
\end{verbatim}
Note that $M_{SUSY}$ is the scale where EWSB conditions are fixed, which is (by
default) defined as $M_{SUSY} = \sqrt{m_{\tilde{t}_1}
  m_{\tilde{t}_2}}$ and is re-calculated at each step in the
iteration. This matches SLHA conventions,%
  \footnote{though in the SLHA papers this scale is named $M_{EWSB}$.}
but in \SOFTSUSY\ there is also a special option to vary this scale by 
setting entry 4 of block \SOFTSUSY\ to give the value of QEWSB, which alters 
$M_{SUSY}$ as described in section A.1 of the MSSM manual \cite{Allanach:2001kg}.

Instead of choosing the default EWSB output parameters (\ref{eq:Z3v_ewsb}-\ref{eq:Z3_ewsb}) 
it is also possible to output the soft scalar Higgs masses.  The EWSB conditions in 
Eqs.~(\ref{eq:mind}-\ref{eq:mins}) will then determine $m_{H_1}^2$, $m_{H_2}^2$ and $m_S^2$ 
when entry $18$ in block \SOFTSUSY\ is set to $1$:
\begin{verbatim}
Block SOFTSUSY
   18   1                    # use soft Higgs masses as EWSB output
\end{verbatim}

In this case, the default EWSB output parameters (\ref{eq:Z3v_ewsb}-\ref{eq:Z3_ewsb}) must be given in the SLHA file or they will be set to zero by default. Since all $\Zv$ parameters are inputs in this case, setting any of them to a non-zero value implies that the point considered belongs to the $\Zv$-NMSSM. In this particular case, entries 23, 24 and 67 in block {\tt EXTPAR} may now be set for a $\Zv$-NMSSM point since they are no longer EWSB outputs.

For the SLHA2 input option, 
the output will also be given in 
SLHA2 format. Such output can be used for
input into other programs which subscribe to the accord, such as
\code{PYTHIA}~\cite{Sjostrand:2007gs} (for
simulating sparticle production and decays at colliders), for example. For
further details on the format of 
the input and output files, see Refs.~\cite{Allanach:2008qq} and
\cite{Skands:2003cj}. 

An alternative input option for \SOFTSUSY\ is to input the parameters via the command-line interface. As of {\tt SOFTSUSY 3.4}, the command line interface of \code{softpoint.x} has
changed, see \code{softpoint.x --help}.  For the NMSSM, the syntax is
\small
\begin{verbatim}
 ./softpoint.x nmssm <susy-breaking-model> [NMSSM flags] [NMSSM parameters] [general options]
\end{verbatim}
\normalsize
where \code{sugra} is the only currently available susy-breaking
model.  The general options are listed in Ref.~\cite{Allanach:2001kg}
%\tablename~\ref{tab:general-cmd-line-options} 
and the NMSSM flags and
parameter options are listed in
\tablename~\ref{tab:nmssm-cmd-line-options}.
\begin{table}[tbh]
  \centering
  \begin{tabular}{ll}
    NMSSM flags & description \\
    \hline
    \code{--lambdaAtMsusy} & input $\lambda$ at scale $M_{SUSY}$ \\
    \hline\\
    NMSSM parameters & description \\
    \hline
    \code{--m0=<value>} & unified soft scalar mass \\
    \code{--m12=<value>} & unified soft gaugino mass \\
    \code{--a0=<value>} & unified trilinear coupling \\
    \code{--tanBeta=<value>} & $\tan\beta$ \\
    \code{--mHd2=<value>} & soft down-type Higgs mass squared $m_{H_1}^2$ \\
    \code{--mHu2=<value>} & soft up-type Higgs mass squared $m_{H_2}^2$ \\
    \code{--mu=<value>} & $\mu$ parameter \\
    \code{--m3SqrOverCosBetaSinBeta=<value>} & $m_3^2/(\cos\beta \sin\beta)$ \\
    \code{--lambda=<value>} & trilinear superpotential coupling $\lambda$ \\
    \code{--kappa=<value>} & trilinear superpotential coupling $\kappa$ \\
    \code{--Alambda=<value>} & trilinear soft coupling $A_\lambda$ \\
    \code{--Akappa=<value>} & trilinear soft coupling $A_\kappa$ \\
    \code{--lambdaS=<value>} & $\lambda \langle S \rangle = \lambda s / \sqrt{2}$ \\
    \code{--xiF=<value>} & linear superpotential coupling $\xi_F$ \\
    \code{--xiS=<value>} & linear soft coupling $\xi_S$ \\
    \code{--muPrime=<value>} & bilinear superpotential coupling $\mu'$ \\
    \code{--mPrimeS2=<value>} & bilinear soft coupling $m_{S}'^2$ \\
    \code{--mS2=<value>} & bilinear soft mass $m_{S}^2$ \\
    \hline
  \end{tabular}
  \caption{NMSSM command line options for \code{softpoint.x}}
  \label{tab:nmssm-cmd-line-options}
\end{table}

\section{Calculating decays with \NMSSMTools\label{sec:decays}}

\SOFTSUSY\ has a compatibility mode which interfaces with \NMSSMTools\
to calculate sparticle decays in the NMSSM.  To enable it, the
user has to first install \NMSSMTools\ and then run the
\code{setup\_nmssmtools.sh} script
\begin{verbatim}
  $ cd /path/to/NMSSMTools/
  $ wget http://www.th.u-psud.fr/NMHDECAY/NMSSMTools_4.1.2.tgz
  $ tar xf NMSSMTools_4.1.2.tgz
  $ cd /path/to/softsusy/
  $ ./setup_nmssmtools.sh \
       --nmssmtools-dir=/path/to/NMSSMTools/NMSSMTools_4.1.2 \
       --compile
\end{verbatim}
The \code{setup\_nmssmtools.sh} script copies \code{nmProcessSpec.f}
and \code{Makefile.nmssmtools} from the \SOFTSUSY\ directory to the
{\tt main/} directory within the \NMSSMTools\ folder.  If the \code{--compile}
flag is provided, \NMSSMTools\ is recompiled.  Afterwards the user can
generate a NMSSM spectrum with \SOFTSUSY\ and use \NMSSMTools\ to
calculate the decays.  The \code{softsusy\_nmssmtools.x} script combines
these two steps:
\begin{verbatim}
  $ ./softsusy_nmssmtools.x leshouches < slhaInput > slhaOutput
\end{verbatim}
Here \code{slhaInput} is an SLHA input file with the SOFTSUSY block
entry $15$ set to $1$.  Additional \NMSSMTools\ specific flags can also
be used with entries $16$ and $17$, which are past to \NMSSMTools\ as
MODSEL blocks $9$ and $10$ respectively, following the \NMSSMTools\
convention.
\begin{verbatim}
   Block SOFTSUSY
      15   1      # NMSSMTools compatible output (default: 0) 
      16   4      # Select Micromegas option for NMSSMTools
                  # (default: 0) 0=no, 1=relic density only
                  # 2=direct detection + relic density, 
                  # 3=indirect detection + relic density
                  # 4=all  
      17   1      # 1:sparticle decays via NMSDECAY (default: 0)
\end{verbatim}
After \code{softsusy\_nmssmtools.x} is called, the following three output
files can be found in the \NMSSMTools\ directory
\code{NMSSMTools\_4.1.2/main/}. The file \code{nmProcessSpec-decay}
contains the sparticle decays in form of SLHA DECAY blocks,
\code{nmProcessSpec-omega} will contain the output from \code{micrOMEGAS} if entry 16 is selected to be non-zero and
\code{nmProcessSpec-spectr} contains the spectrum calculated by
\NMSSMTools.

\section{Class Structure\label{sec:objects}}

We now go on to sketch the NMSSM class hierarchy.  Only methods and
data which are deemed of possible importance for prospective users are
mentioned here, but there are many others within the program itself.

\subsection{General structure}

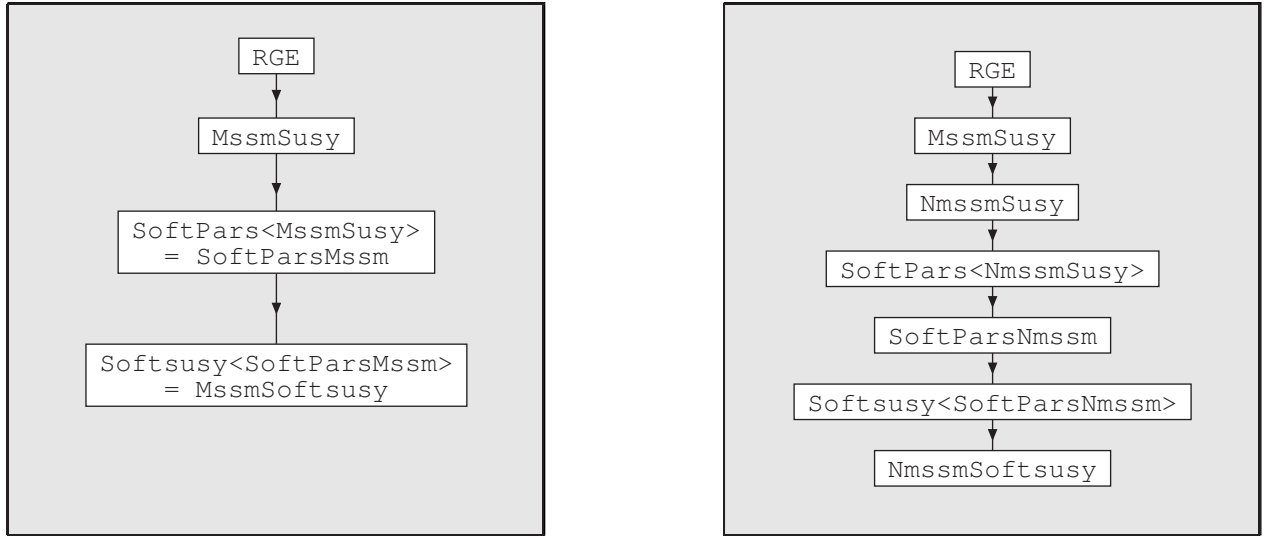
\begin{figure}
  \begin{center}
    \begin{picture}(200,200)
      \GBox(200,200)(0,0){0.9}
      \ArrowLine(100,180)(100,150)
      \ArrowLine(100,150)(100,110)
      \ArrowLine(100,110)(100,60)
      \SetPFont{Teletype}{10}
      \put(0,0){\framebox(200,200){}}
      \BText(100,180){RGE}
      \BText(100,150){MssmSusy}
      \B2Text(100,110){SoftPars<MssmSusy>}{= SoftParsMssm}
      \B2Text(100,60){Softsusy<SoftParsMssm>}{= MssmSoftsusy}
    \end{picture}\hfill
    \begin{picture}(200,200)
      \GBox(200,200)(0,0){0.9}
      \ArrowLine(100,175)(100,150)
      \ArrowLine(100,150)(100,125)
      \ArrowLine(100,125)(100,100)
      \ArrowLine(100,100)(100,75)
      \ArrowLine(100,75)(100,50)
      \ArrowLine(100,50)(100,25)
      \SetPFont{Teletype}{10}
      \put(0,0){\framebox(200,200){}}
      \BText(100,175){RGE}
      \BText(100,150){MssmSusy}
      \BText(100,125){NmssmSusy}
      \BText(100,100){SoftPars<NmssmSusy>}
      \BText(100,75){SoftParsNmssm}
      \BText(100,50){Softsusy<SoftParsNmssm>}
      \BText(100,25){NmssmSoftsusy}
    \end{picture}
    \caption{\label{fig:objstruc} Heuristic high-level class
      structure of \SOFTSUSY. Inheritance is displayed by the
      arrows and {\tt typedef}s are displayed by the equals signs.}
  \end{center}
\end{figure}

To implement the NMSSM (and other non-minimal supersymmetric models),
the \SOFTSUSY~class hierarchy was generalized with the following
requirements in mind:
\begin{itemize}
\item The class of supersymmetric parameters (gauge couplings,
  superpotential parameters and VEVs), whose beta functions are
  independent of soft-breaking parameters, should be at the top of the
  class hierarchy.  This makes them usable independently of the
  soft-breaking parameters, for example during the initial guess.

\item One should be able to reuse as much MSSM code as possible, for
  example by inheriting from existing MSSM classes.
\end{itemize}

The above requirements were implemented by the following changes:
\begin{enumerate}
\item The class of the soft breaking MSSM parameters and their beta
  functions was converted into the class template \code{SoftPars<Susy>}.
  The template parameter represents the class of supersymmetric
  parameters, from which \code{SoftPars<Susy>} inherits.  The class
  which contains \emph{all} MSSM parameters and beta functions,
  \code{SoftParsMssm}, was made a typedef for
  \code{SoftPars<MssmSusy>}, where \code{MssmSusy} is the class that
  contains the supersymmetric MSSM parameters and beta functions.
\begin{verbatim}
template <class Susy>
class SoftPars : public Susy {
   // implementation of soft breaking MSSM parameters
   // and their beta functions
};

typedef SoftPars<MssmSusy> SoftParsMssm;
\end{verbatim}
  This approach makes it possible to have a class of soft breaking
  MSSM parameters but with a different set of supersymmetric
  parameters.  This mechanism is used in the NMSSM, see
  Section~\ref{nmssmsoftpars}. 

\item The class which organises the MSSM mass spectrum calculation was
  converted into the class template \code{Softsusy<SoftPars>}.  The
  template parameter represents the class of all model parameters and
  beta functions, from which \code{Softsusy<SoftPars>} inherits.
  \code{MssmSoftsusy} was made a \code{typedef} for
  \code{Softsusy<SoftParsMssm>}.
\begin{verbatim}
template <class SoftPars>
class Softsusy : public SoftPars {
   // organisation of MSSM mass spectrum calculation
   // using model parameters in SoftPars
};

typedef Softsusy<SoftParsMssm> MssmSoftsusy;
\end{verbatim}
  This approach makes it possible to have a MSSM spectrum calculation
  class but with an arbitrary set of model parameters.  This mechanism
  is used in the NMSSM, see~\ref{nmssmsoftsusy}.
\end{enumerate}

\subsection{\code{NmssmSusy}~class}
\label{nmssmsusy}

The class of supersymmetric NMSSM parameters and beta functions,
\code{NmssmSusy}, inherits from \code{MssmSusy} to reuse the MSSM
parameters and beta functions, see \figurename~\ref{fig:objstruc}.  It
adds data members and access methods for the new supersymmetric NMSSM
parameters, which can be found in \tablename~\ref{tab:nmssmsusy}.
\begin{table}
  \centering
  \begin{tabular}{lll}
    data variable & & methods \\\hline
    \code{\small double lambda, kappa} & trilinear superpotential &
    \code{\small displayLambda}
    \\
    $\lambda$, $\kappa$ & couplings & \code{\small displayKappa}
    \\\hline
    \code{\small double mupr} & bilinear superpotential &
    \code{\small displayMupr}
    \\
    $\mu'$ & coupling &
    \\\hline
    \code{\small double xiF} & linear superpotential &
    \code{\small displayXiF}
    \\
    $\xi_F$ & coupling &
    \\\hline
    \code{\small double sVEV} & VEV of singlet field &
    \code{\small displaySVEV}
    \\
    $s$ & &
    \\\hline
    \normalsize
  \end{tabular}
  \caption{\code{NmssmSusy} class data and accessor methods
    \label{tab:nmssmsusy}}
\end{table}

\subsection{\code{SoftParsNmssm}~class}
\label{nmssmsoftpars}

To implement the class of soft-breaking NMSSM parameters,
\code{SoftParsNmssm}, the \code{SoftPars<Susy>} template is
instantiated using \code{NmssmSusy} as template parameter.  Thereby
one obtains the class of MSSM soft-breaking beta functions, using
supersymmetric NMSSM parameters.  \code{SoftParsNmssm} then inherits
from \code{SoftPars<NmssmSusy>} to add extra NMSSM contributions to
the soft-breaking beta functions:
\begin{verbatim}
class NmssmSusy : public MssmSusy {
   // implement supersymmetric NMSSM parameter beta functions
   // by reusing MSSM ones
};

class SoftParsNmssm : public SoftPars<NmssmSusy> {
   // implement soft-breaking NMSSM parameter beta functions
   // by reusing MSSM ones
};
\end{verbatim}
Furthermore, \code{SoftParsNmssm} adds new soft-breaking NMSSM data
members and access methods, which are listed in
\tablename~\ref{tab:nmssmsoftpars}.
\begin{table}
  \centering
  \begin{tabular}{lll}
    data variable & & methods \\\hline
    \code{\small double alambda, akappa} & trilinear soft &
    \code{\small displayTrialambda}
    \\
    $a_\lambda$, $a_\kappa$ & parameters & \code{\small displayTriakappa}
    \\\hline
    $A_\lambda$ & $a_\lambda / \lambda$ & \code{\small displaySoftAlambda}
    \\
    $A_\kappa$ & $a_\kappa / \kappa$ & \code{\small displaySoftAkappa}
    \\\hline
    \code{\small double mSpsq} & bilinear soft &
    \code{\small displayMspSquared}
    \\
    $m_{S}'^2$ & parameters &
    \\\hline
    \code{\small double mSsq} & soft scalar mass &
    \code{\small displayMsSquared}
    \\
    $m_S^2$ & &
    \\\hline
    \code{\small double xiS} & linear soft &
    \code{\small displayXiS}
    \\
    $\xi_S$ & parameters &
    \\\hline
    \normalsize
  \end{tabular}
  \caption{\code{SoftParsNmssm} class data and accessor methods
    \label{tab:nmssmsoftpars}}
\end{table}

\subsection{\code{NmssmSoftsusy}~class}
\label{nmssmsoftsusy}

To create the NMSSM spectrum calculation class, \code{NmssmSoftsusy},
the \code{Softsusy<SoftPars>} template class is instantiated using
\code{SoftParsNmssm} as template parameter.  Thereby one obtains an
NMSSM spectrum calculator, which uses NMSSM parameters and beta
functions.  \code{NmssmSoftsusy} then inherits from
\code{Softsusy<SoftParsNmssm>} and overwrites MSSM functions to
account for the extra NMSSM particles:
\begin{verbatim}
class NmssmSoftsusy : public Softsusy<SoftParsNmssm> {
   // organise NMSSM spectrum calculation reusing MSSM functions
};
\end{verbatim}
To implement the NMSSM pole masses and mixing matrices, the
\code{sPhysical} structure had to be generalized, as in
\tablename~\ref{tab:sphys}.
\begin{table}
  \centering
  \begin{tabular}{ll}
    data variable & description \\ \hline
    \code{DoubleVector mh0,mA0} & vectors of neutral Higgs masses $m_{h^0_{1\ldots n}}, m_{A^0_{1\ldots m}}$\\
    & (MSSM: $n=2, m=1$, NMSSM: $n=3, m=2$) \\
    \code{double mHpm} & charged Higgs mass $m_{H^\pm}$ \\
    \code{DoubleVector msnu} & vector of $m_{{\tilde \nu}_{i=1 \ldots 3}}$ masses \\
    \code{DoubleVector mch,mneut} & vectors of $m_{{\chi^\pm}_{i=1 \ldots 2}}$, 
    $m_{{\chi^0}_{i=1 \ldots n}}$ respectively \\
    & (MSSM: $n=4$, NMSSM: $n=5$) \\
    \code{double mGluino} & gluino mass $m_{\tilde g}$ \\
    \code{DoubleMatrix mixNeut} & orthogonal neutralino mixing matrix $O$\\
    & (MSSM: 4 by 4, NMSSM: 5 by 5)\\
    \code{double thetaL, thetaR} & $\theta_{L, R}$ chargino mixing angles \\
    \code{double thetat, thetab} & $\theta_{t,b}$ sparticle mixing angles \\
    \code{double thetatau} & $\theta_{\tau}$ sparticle mixing angle \\
    \code{double thetaH} & CP-even Higgs mixing angle $\alpha$ in the MSSM \\
    \code{double thetaA0} & CP-odd Higgs mixing angle $\theta_{A^0}$ in the NMSSM \\
    \code{DoubleMatrix mixh0} & orthogonal CP-even Higgs mixing matrix $R$ in the NMSSM \\
    \code{DoubleMatrix mu, md, me} & (2 by 3) matrices of up squark, down squark
    and\\
    &  charged slepton masses \\
  \end{tabular}
  \caption{\label{tab:sphys}\code{sPhysical} structure. Masses are pole
    masses, and stored in units of GeV. Mixing angles are in radian
    units.}
\end{table}

\section{Renormalization Group Equations for the NMSSM}\label{sec:RGEs}
In this section, we present the components of the one- and two-loop renormalization group equations (RGEs) which belong exclusively to the NMSSM.  Our expressions have been derived in the $\overline{\mbox{DR}}$ scheme from existing results \cite{MV94,Yam94} for general SUSY gauge theories. The complete RGEs are then obtained by combing the expressions below with those for the MSSM \cite{MV94}.

\subsection{Yukawa Couplings}
For $t = \ln Q$, the trilinear superpotential parameter $Y^{ijk}$ evolves 
according to the general expression \cite{MV94}
\begin{equation}
\dt Y^{ijk} = Y^{ijp}\Gamma_p^k + Y^{kjp}\Gamma_p^i + Y^{ikp}\Gamma_p^j\,,
\label{eq:Yuk rges}
\end{equation}
where 
\begin{equation}
\Gamma_i^j = \frac{1}{16\pi^2}\gamma_i^{(1)j} 
+ \frac{1}{(16\pi^2)^2}\gamma_{i}^{(2)j}\,, 
\end{equation}
and $\gamma^{(1,2)j}_i$ are the one- and two-loop anomalous dimensions 
respectively.  Note that the $3\times 3$ Yukawa matrices $Y_{U,D,E}$ are 
obtained by identifying indices in Eq.~(\ref{eq:Yuk rges}) with the relevant chiral superfields in the superpotential.%
\footnote{For example, for $k=H_2$ we have $Y^{ijH_2}\equiv (Y_U)^{ij}$.}  

At one-loop order, the only addition to the MSSM expressions 
\cite{MV94} for the $Y_{U,D,E}$ RGEs is the inclusion of $\lamsq$ terms 
which originate from the Higgs anomalous dimensions
\begin{align}
\left.\gamma^{(1) H_1}_{H_1}\right|_\lambda = \lamsq 
\quad \mbox{and} \quad \left.\gamma^{(1) H_2}_{H_2} \right|_\lambda = \lamsq\,.
\end{align}
At two-loop order, all the gauge-Yukawa contributions from $\lambda$ cancel for 
each $\gamma_i^{(2)j}$, so the additional contributions arising in the NMSSM are 
simply given by
\begin{align}
\left.\gamma_{L_i}^{(2)L_j}\right|_\lambda &= -\lamsq (Y_E Y_E^\dagger)_i^j\,, \\
\left.\gamma_{E_i}^{(2)E_j}\right|_\lambda &= -2\lamsq (Y_E^\dagger Y_E)_i^j\,, \\
\left.\gamma_{Q_i}^{(2)Q_j}\right|_\lambda &= -\lamsq (Y_U Y_U^\dagger)_i^j 
- \lamsq (Y_D Y_D^\dagger)_i^j\,, \\
\left.\gamma_{D_i}^{(2)D_j}\right|_\lambda &= -2\lamsq (Y_D^\dagger Y_D)_i^j\,, \\
\left.\gamma_{U_i}^{(2)U_j}\right|_\lambda &= -2\lamsq (Y_U^\dagger Y_U)_i^j \,, \\
\left.\gamma_{H_1}^{(2)H_1}\right|_\lambda &= -3\lambda^4 -2\lamsq\kapsq 
- 3\lamsq \tr(Y_U Y_U^\dagger)\,, \\
\left.\gamma_{H_2}^{(2)H_2}\right|_\lambda &= -3\lambda^4 -2\lamsq\kapsq 
- 3\lamsq \tr(Y_D Y_D^\dagger) - \lamsq \tr(Y_E Y_E^\dagger)\,.
\end{align}

In a similar manner, the RGEs for $\lambda$ and $\kappa$ are obtained from Eq.~(\ref{eq:Yuk rges}), with
\begin{align}
\dt\lambda &= \lambda (\Gamma^{H_1}_{H_1} + \Gamma^{H_2}_{H_2} + \Gamma^S_S)\,, \\
\dt\kappa &= 3\kappa \Gamma_S^S\,,
\end{align}
where the one- and two-loop expressions for the singlet anomalous dimension 
are given by
\begin{align}
\gamma_S^{(1)S} &= 2\lamsq + 2\kapsq\,,\\
\gamma_S^{(2)S} &= -4\lambda^4 - 8\kappa^4 - 8 \kapsq\lamsq 
- 6\lamsq \tr(Y_UY_U^\dagger) - 6\lamsq \tr(Y_UY_U^\dagger) 
- 2\lamsq \tr(Y_EY_E^\dagger) + \tfrac{6}{5}g_1^2\lamsq + 6g_2^2\lamsq\,.
\end{align}

\subsection{Gauge Couplings}
In the NMSSM, the one-loop RGEs for the gauge couplings $g_a$ are 
identical to those for the MSSM.  At two-loop order however, the $\lambda$ 
coupling appears through the term
\begin{equation}
\dt g_a \ni - \frac{g_a^3}{(16\pi^2)^2} Y_{ijk}Y^{ijk} C_a(k)/d(G_a)\,,  
\label{eqn:dg}
\end{equation}
where $d(G_a)$ is the dimension of the adjoint representation of gauge group 
$G_a$.  The result is
\begin{equation}
\left. \dt g_a\right|_\lambda = -\frac{g_a^3}{(16\pi^2)^2}\lamsq \Lambda_a^{(2)}\,,
\qquad \Lambda_a^{(2)} = (\tfrac{6}{5},2,0)\,,
\label{eq:dg}
\end{equation}
where we have taken into account the additional factor of 2 which arises from 
tracing over $SU(2)$ group indices in Eq.~(\ref{eqn:dg}).

\subsection{Gaugino Mass Parameters}
As for the gauge couplings above, we need only consider the addition of the 
$\lambda^2$ terms arising from
\begin{equation}
\dt M_a \ni \frac{2g_a^2}{(16\pi^2)^2} 
\frac{(T_A^{ijk} - M_a Y^{ijk}) Y_{ijk}C_a(k)}{d(G_a)}\,,
\label{eqn:dM}
\end{equation}
where $T_A^{ijk}$ is a trilinear soft SUSY-breaking parameter. By evaluating the summations in Eq.~(\ref{eqn:dM}), we find
\begin{equation}
\left. \dt M_a\right|_{\lambda} = \frac{2g_a^2}{(16\pi^2)^2} (\lambda a_\lambda - \lambda^2 M_a)\Lambda^{(2)}_a\,,
\end{equation}
with $\Lambda_a^{(2)}$ as given in (\ref{eq:dg}).

\subsection{$\mu$ Parameters}
The general expression \cite{MV94,Yam94} for the SUSY-conserving bilinear terms is given by  
\begin{equation}
\dt \mu^{ij} = \mu^{ip}\Gamma_p^j + \mu^{jp}\Gamma_p^i\,,
\end{equation}
from which we obtain
\begin{align}
\dt\mu &= \mu (\Gamma_{H_1}^{H_1} + \Gamma_{H_2}^{H_2}), \notag \\
\dt\mu' &= 2\mu' \Gamma_S^S
\end{align}

\subsection{Trilinear Couplings}
If we denote $T_A^{ijk}$ as a soft SUSY-breaking trilinear, then the evolution at two-loop is given by 
\begin{equation}
\dt T_A^{ijk} = \frac{1}{16\pi^2} \left[\beta_{T_A}^{(1)}\right]^{ijk} 
+ \frac{1}{(16\pi^2)^2} \left[\beta_{T_A}^{(2)}\right]^{ijk}\,,
\end{equation}
where the explicit expressions for the $\beta$ functions can be found in \cite{MV94}.  
For $T = U,D,E$, the $\lambda$ contribution to the one-loop $\beta$ function 
arises from the following factor
\begin{equation}
\left[\beta_{T_A}^{(1)}\right]^{ij} \ni \tfrac{1}{2} (T_A)^{ij} 
Y_{H_\alpha mn}Y^{mnH_\alpha} + (Y_x)^{ij} Y_{H_\alpha mn}T_A^{mnH_\alpha}\,,
\end{equation}
where there is {\it no summation} over $\alpha = 1,2$, with the index 
determined by the choice of $T$ (e.g.\ if $T=U$ then $\alpha = 2$). Expanding 
the indices leads to
\begin{equation}
\left.\left[\beta_{T_A}^{(1)}\right]^{ij}\right|_\lambda = (T_A)^{ij}\lambda^2 
+ (Y_x)^{ij} 2\lambda a_\lambda\,. \label{eqn: beta hx}
\end{equation}

The two-loop expressions involve a large number of summations so to minimize the proliferation of generation indices we choose to express our results in terms of $3\times 3$ matrices:
\begin{align}
\left.\beta_{ U_A}^{(2)}\right|_\lambda =& -\lamsq U_A\big[ 3\lamsq + 2\kapsq 
+ 3\tr( Y_D Y_D^\dagger) + \tr( Y_E Y_E^\dagger) \big] 
-\lamsq\big[ 5 Y_U Y_U^\dagger U_A + 4 U_A Y_U^\dagger Y_U +  Y_D Y_D^\dagger U_A 
+ 2 D_A Y_D^\dagger Y_U \big] \notag \\
&-2\lambda  a_\lambda Y_U\big[ 3\lamsq + 2\kapsq + 3\tr( Y_D Y_D^\dagger) 
+ \tr( Y_E Y_E^\dagger) \big] -2\lamsq Y_U\big[ 3\lambda  a_\lambda 
+ 2\kappa a_\kappa + 3\tr( D_A Y_D^\dagger) + \tr( E_A Y_E^\dagger) \big] \notag \\
&-2\lambda  a_\lambda\big[ 3 Y_U Y_U^\dagger Y_U +  Y_D Y_D^\dagger Y_U \big]\,, \\
\left.\beta_{ D_A}^{(2)}\right|_\lambda =& -\lamsq D_A\big[ 3\lamsq + 2\kapsq 
+ 3\tr( Y_U Y_U^\dagger) \big] 
-\lamsq\big[ 5 Y_D Y_D^\dagger D_A + 4 D_A Y_D^\dagger Y_D + 2 U_A Y_U^\dagger Y_D 
+  Y_U Y_U^\dagger D_A \big] \notag \\
&-2\lambda  a_\lambda  Y_D\big[ 3\lamsq + 2\kapsq + 3\tr( Y_U Y_U^\dagger) \big]
-2\lamsq Y_D\big[ 3\lambda  a_\lambda + 2\kappa a_\kappa 
+ 3\tr( U_A Y_U^\dagger) \big] -2\lambda  a_\lambda\big[ 3 Y_D Y_D^\dagger Y_D 
+  Y_U Y_U^\dagger Y_D\big]\,,  \\
\left.\beta_{ E_A}^{(2)}\right|_\lambda =& -\lamsq  E_A\big[ 3\lamsq + 2\kapsq 
+ 3\tr( Y_U Y_U^\dagger) \big] 
-\lamsq\big[ 5 Y_E Y_E^\dagger E_A + 4 E_A Y_E^\dagger Y_E \big] 
-2\lambda  a_\lambda  Y_E\big[ 3\lamsq + 2\kapsq + 3\tr( Y_U Y_U^\dagger) \notag \\
&-2\lamsq Y_E\big[ 3\lambda a_\lambda + 2\kappa a_\kappa + 3\tr( U_A Y_U^\dagger) 
\big] -6\lambda a_\lambda  Y_E Y_E^\dagger Y_E\,.
\end{align}

For $a_\lambda$, the one-loop $\beta$ function reads in full
\begin{align}
\beta_{a_\lambda}^{(1)} =&\, \tfrac{1}{2}a_\lambda (Y_{H_1mn}Y^{mnH_1} + Y_{H_2mn}Y^{mnH_2}
 + Y_{Smn}Y^{mnS}) + \lambda (Y_{H_1mn}T_A^{mnH_1} + Y_{H_2mn}T_A^{mnH_2} 
+ Y_{Smn}T_A^{mnS}) \notag \\
&- 4\sum_{a=1,2,3} (a_\lambda - 2M_a\lambda)g_a^2C_a(H)\,,
\end{align}
from which the various sums immediately yield
\begin{align}
\beta_{a_\lambda}^{(1)} =&\, a_\lambda[3\tr( Y_U  Y_U^\dagger) + 3\tr( Y_D  Y_D^\dagger) 
+ \tr( Y_E  Y_E^\dagger) + 12\lamsq + 2\kapsq - \tfrac{3}{5}g_1^2 - 3g_2^2] 
\notag \\
&+ \lambda[6\tr( U_A  Y_U^\dagger) + 6\tr( D_A  Y_D^\dagger) + 2\tr( E_A  Y_E^\dagger)
+ 4a_\kappa\kappa + \tfrac{6}{5}g_1^2M_1 + 6g_2^2M_2]\,.
\end{align}
The two-loop expression is given by
\begin{align}
\beta_{a_\lambda}^{(2)} =&-50\lambda^4 a_\lambda 
- 36\lambda\tr(U_A Y_U^\dagger Y_U Y_U^\dagger) 
- 36\lambda\tr(D_A Y_D^\dagger Y_D Y_D^\dagger)
- 12\lambda\tr(E_A Y_E^\dagger Y_E Y_E^\dagger) 
- 9a_\lambda\tr( Y_U Y_U^\dagger Y_U Y_U^\dagger) \notag \\
&- 9a_\lambda\tr( Y_D Y_D^\dagger Y_D Y_D^\dagger) 
- 3a_\lambda\tr( Y_E Y_E^\dagger Y_E Y_E^\dagger) - 8\kappa^4 a_\lambda 
- 32\lambda\kappa^3 a_\kappa - 12\lamsq\kapsq a_\lambda \notag \\
&-18\lambda^3\big[ (\Alam)\tr( Y_U Y_U^\dagger) + \tr(U_A Y_U^\dagger)\big] 
- 18\lambda^3\big[(\Alam)\tr( Y_D Y_D^\dagger) + \tr(D_A Y_D^\dagger)\big] \notag \\
&-6\lambda^3\big[(\Alam)\tr( Y_E Y_E^\dagger) + \tr(E_A Y_E^\dagger)\big] 
-24\lambda^3\kapsq\big[ (\Alam) + (\Akap)\big]
-12\lambda\big[ \tr(U_A Y_U^\dagger Y_D Y_D^\dagger) 
+ \tr(D_A Y_D^\dagger Y_U Y_U^\dagger) \big] \notag \\
&-3\lamsq a_\lambda\big[ 3\tr( Y_U Y_U^\dagger) + 3\tr( Y_D Y_D^\dagger) 
+ \tr( Y_E Y_E^\dagger) \big] -6a_\lambda\tr( Y_U Y_U^\dagger Y_D Y_D^\dagger) 
+ \tfrac{12}{5}g_1^2\lamsq\big[ \tfrac{3}{2}a_\lambda - \lambda M_1\big] \notag \\
&+\tfrac{8}{5}g_1^2\lambda\big[ \tr(U_A Y_U^\dagger) - M_1\tr( Y_U Y_U^\dagger) \big]
-\tfrac{4}{5}g_1^2\lambda\big[ \tr(D_A Y_D^\dagger) - M_1\tr( Y_D Y_D^\dagger)\big] 
+\tfrac{12}{5}g_1^2\lambda\big[\tr(E_A Y_E^\dagger) - M_1\tr( Y_E Y_E^\dagger)\big] 
\notag \\
&+\tfrac{2}{5}g_1^2 a_\lambda\big[ 2\tr( Y_U Y_U^\dagger) - \tr( Y_D Y_D^\dagger) 
+ 3\tr( Y_E Y_e^\dagger) \big] + 12g_2^2\lamsq\big[ \tfrac{3}{2}a_\lambda 
- \lambda M_2\big] \notag \\
&+32g_3^2\lambda\big[ \tr(U_A Y_U^\dagger) - M_3\tr( Y_U Y_U^\dagger) \big]
+32g_3^2\lambda\big[\tr(D_A Y_D^\dagger) - M_3\tr( Y_D Y_D^\dagger) \big] 
+ 16g_3^2 a_\lambda\big[ \tr( Y_U Y_U^\dagger) + \tr( Y_D Y_D^\dagger) \big] \notag \\
&+ \tfrac{1}{50}g_1^4\lambda\big[ 207(\Alam) - 828M_1\big] 
+ \tfrac{1}{2}g_2^4\lambda\big[ 15(\Alam) - 60M_2\big]
+\tfrac{9}{5}g_1^2 g_2^2\lambda\big[ (\Alam) - 2(M_1 + M_2)\big] \,.
\end{align}

For $a_\kappa$, the one-loop calculation is similar to that of $a_\lambda$, with 
the result
\begin{equation}
\beta_{a_\kappa}^{(1)} = 18a_\kappa\kappa^2 + 12a_\lambda\kappa\lambda 
+ 6a_\kappa\lamsq\,.
\end{equation}
At two-loop we have
\begin{align}
\beta_{a_\kappa}^{(2)} =& -120\kappa^4 a_\kappa - 12\lambda^4 a_\kappa 
- 48\lambda^3\kappa a_\lambda - 48\lamsq\kappa^3\big[ (\Alam) + (\Akap)\big]
-24\lamsq\kapsq a_\kappa \notag \\
&-36\lamsq\kappa\big[ \tr( U_A Y_u^\dagger) + (\Alam)\tr( Y_u Y_u^\dagger)\big]
-36\lamsq\kappa\big[ \tr( D_A Y_d^\dagger) + (\Alam)\tr( Y_d Y_d^\dagger)\big] 
\notag \\
&-12\lamsq\kappa\big[ \tr( E_A Y_e^\dagger) + (\Alam)\tr( Y_e Y_e^\dagger)\big]
-6\lamsq a_\kappa\big[ 3\tr( Y_u Y_u^\dagger) + 3\tr( Y_d Y_d^\dagger) 
+ \tr( Y_e Y_e^\dagger) \big] \notag \\
&+ \tfrac{36}{5}g_1^2\lamsq\kappa\big[ (\Alam) + \tfrac{1}{2}(\Akap) - M_1\big] 
+ 36g_2^2\lamsq \kappa\big[ (\Alam) + \tfrac{1}{2}(\Akap) - M_2 \big]\,.
\end{align}

\subsection{Higgs Masses}
To determine the $\lambda$ and $\kappa$ contributions to the Higgs masses, it is useful to define \cite{Ellwanger:2009dp} the following quantities 
\begin{align}
\mlamsq =& \mhdsq + \mhusq + \mssq + a_\lambda^2/\lamsq\,, \notag \\
\mkapsq =& 3\mssq + a_\kappa^2/\kapsq\,, \notag \\
\Musq =& \tr(\mqsq Y_u Y_u^\dagger) + \tr( Y_u\musq Y_u^\dagger)
+ \mhusq\tr( Y_u Y_u^\dagger) + \tr( U_A U_A^\dagger)\,, \notag \\
\Mdsq =& \tr(\mqsq Y_d Y_d^\dagger) + \tr( Y_d\mdsq Y_d^\dagger) 
+ \mhdsq\tr( Y_d Y_d^\dagger) + \tr( D_A D_A^\dagger)\,, \notag \\
\Mesq =& \tr(\mlsq Y_e Y_e^\dagger) + \tr( Y_e\mesq Y_e^\dagger) 
+ \mhdsq\tr( Y_e Y_e^\dagger) + \tr( E_A E_A^\dagger)\,. \notag \\
\end{align}
Both the up- and down-type Higgs masses $m_{H_2}$ and $m_{H_1}$ receive the same 
$\lambda$ contribution at one-loop order, 
\begin{equation}
\left.\beta_{m_{H_\alpha}^2}^{(1)}\right|_\lambda = 2\lamsq \mlamsq\,, \qquad \alpha = 1,2\,.
\end{equation}

The two-loop expressions for $m_{H_2}^2$ are 
\begin{align}
\left.\beta_{m_{H_2}^2}^{(2)}\right|_\lambda =& -12\lambda^4\big\{ M_\lambda^2 + (\Alam)^2\big\} 
- 6\lamsq\big\{ \Mdsq + M_\lambda^2\tr( Y_d Y_d^\dagger) + 2(a_\lambda/\lambda)\tr( D_A Y_d^\dagger) \big\} \notag \\
&-2\lamsq\big\{ \Mesq + M_\lambda^2\tr( Y_e Y_e^\dagger) + 2(a_\lambda/\lambda)\tr( E_A Y_e^\dagger)\big\}
-4\lamsq\kapsq\big\{ M_\lambda^2 + M_\kappa^2 + 2(a_\lambda/\lambda)(a_\kappa/\kappa) \big\}  \notag \\
&+ \tfrac{6}{5}g_1^2\lamsq(\mhdsq-\mhusq) \,,
\end{align}
with a similar result for $m_{H_1}^2$,
\begin{align}
\left.\beta_{m_{H_1}^2}^{(2)}\right|_\lambda =& -12\lambda^4\big\{M_\lambda^2 + (\Alam)^2\big\} 
- 6\lamsq\big\{ \Musq + M_\lambda^2\tr( Y_u Y_u^\dagger) + 2(a_\lambda/\lambda)\tr( U_A Y_u^\dagger) \big\} \notag \\
&-4\lamsq\kapsq\big\{ M_\lambda^2 + M_\kappa^2 + 2(a_\lambda/\lambda)(a_\kappa/\kappa) \big\} 
- \tfrac{6}{5}g_1^2\lamsq(\mhdsq-\mhusq)\,.
\end{align}

For the singlet mass $m_S$, the one-loop result is
\begin{equation}
\beta_{m_S^2}^{(1)} = Y_{Spq}Y^{pqS}m_S^2 + 2Y_{Spq}Y^{Spr}(m^2)^q_r + h_{Spq}h^{Spq}\,, \label{eqn:beta mSsq}
\end{equation}
where 
\begin{align}
Y_{Spq}Y^{Spr}(m^2)^q_r &= 2\lamsq(\mhdsq + \mhusq) + 4\kapsq m_S^2\,, \notag \\
h_{Spq}h^{Spq} &= 4a_\lambda^2 + 4a_\kappa^2\,,
\end{align}
and thus Eq.~(\ref{eqn:beta mSsq}) becomes
\begin{equation}
\beta_{m_S^2}^{(1)} = 4\lamsq \mlamsq + 4\kapsq \mkapsq\,.
\end{equation}
At two-loop we get
\begin{align}
\beta_{m_S^2}^{(2)} =& -16\lambda^4\big\{ \mlamsq + (\Alam)^2\big\} - 32\kappa^4\big\{ \mkapsq + (\Akap)^2\big\}
-12\lamsq\big\{ \mlamsq\tr( Y_u Y_u^\dagger) + \Musq + 2(\Alam)\tr( U_A Y_u^\dagger)\big\} \notag \\
&- 12\lamsq\big\{ \mlamsq\tr( Y_d Y_d^\dagger) + \Mdsq + 2(\Alam)\tr( D_A Y_d^\dagger) \big\} 
- 4\lamsq\big\{ \mlamsq\tr( Y_e Y_e^\dagger) + \Mesq + 2(\Alam)\tr( E_A Y_e^\dagger) \big\} \notag \\
&- 16\lamsq\kapsq\big\{ \mlamsq + \mkapsq + 2(\Alam)(\Akap) \big\}
+ \tfrac{12}{5}g_1^2\lamsq \big\{ \mlamsq - 2M_1[(\Alam) - M_1] \big\} \notag \\
&+ 12g_2^2\lamsq\big\{ \mlamsq -2M_2[(\Alam) - M_2]\big\}
\end{align}

\subsection{Squark and Slepton Masses}
The squark and slepton masses only receive contributions from $\lambda,\kappa$ at two-loop order.  The results are listed below, where $\mathbf{1}$ is a $3\times 3$ unit matrix.
\begin{align}
\left.\beta_{\mqsq}^{(2)}\right|_\lambda =& -\lamsq \big\{ 2 Y_u^\dagger\musq Y_u + \mqsq Y_u Y_u^\dagger +  Y_u Y_u^\dagger\mqsq \notag 
+ 2\mhusq Y_u Y_u^\dagger + 2 U_A U_A^\dagger + 2\mlamsq Y_u Y_u^\dagger 
+ 2a_\lambda/\lambda(  Y_u U_A^\dagger +  U_A Y_u^\dagger) \big\} \notag \\
&-\lamsq\big\{ 2 Y_d^\dagger\mdsq Y_d + \mqsq Y_d Y_d^\dagger +  Y_d Y_d^\dagger\mqsq 
+ 2\mhdsq Y_d Y_d^\dagger + 2 D_A D_A^\dagger + 2\mlamsq Y_d Y_d^\dagger 
+ 2a_\lambda/\lambda(  Y_d D_A^\dagger +  D_A Y_d^\dagger) \big\} \notag \\
&+ \tfrac{2}{5}g_1^2\lamsq(\mhdsq - \mhusq)\mathbf{1}\,,
\end{align}
\begin{align}
\left.\beta_{\musq}^{(2)}\right|_\lambda =& -2\lamsq \big\{ 2 Y_u^\dagger\mqsq Y_u + \musq Y_u^\dagger Y_u +  Y_u^\dagger Y_u\musq 
+ 2\mhusq Y_u^\dagger Y_u + 2 U_A^\dagger U_A + 2\mlamsq Y_u^\dagger Y_u 
+ 2a_\lambda/\lambda(  Y_u^\dagger U_A +  U_A^\dagger Y_u) \big\}  \notag \\
&- \tfrac{8}{5}g_1^2\lamsq(\mhdsq - \mhusq)\mathbf{1}\,,
\end{align}
\begin{align}
\left.\beta_{\mdsq}^{(2)}\right|_\lambda =& -2\lamsq \big\{ 2 Y_d^\dagger\mqsq Y_d + \mdsq Y_d^\dagger Y_d +  Y_d^\dagger Y_d\mdsq 
+ 2\mhdsq Y_d^\dagger Y_d + 2 D_A^\dagger D_A + 2\mlamsq Y_d^\dagger Y_d 
+ 2a_\lambda/\lambda(  Y_d^\dagger D_A +  D_A^\dagger Y_d) \big\} \notag \\
&+ \tfrac{4}{5}g_1^2\lamsq(\mhdsq - \mhusq)\mathbf{1}\,,
\end{align}
\begin{align}
\left.\beta_{\mlsq}^{(2)}\right|_\lambda =& -\lamsq \big\{ 2 Y_e^\dagger\mesq Y_e + \mlsq Y_e Y_e^\dagger +  Y_e Y_e^\dagger\mlsq 
+ 2\mhdsq Y_e Y_e^\dagger + 2 E_A E_A^\dagger + 2\mlamsq Y_e Y_e^\dagger 
+ 2a_\lambda/\lambda(  Y_e E_A^\dagger +  E_A Y_e^\dagger) \big\} \notag \\
&- \tfrac{6}{5}g_1^2\lamsq(\mhdsq - \mhusq)\mathbf{1}\,,
\end{align}
\begin{align}
\left.\beta_{\mesq}^{(2)}\right|_\lambda =& -2\lamsq \big\{ 2 Y_e^\dagger\mlsq Y_e + \mesq Y_e^\dagger Y_e +  Y_e^\dagger Y_e\mesq 
+ 2\mhdsq Y_e^\dagger Y_e + 2 E_A^\dagger E_A + 2\mlamsq Y_e^\dagger Y_e
+ 2a_\lambda/\lambda(  Y_e^\dagger E_A +  E_A^\dagger Y_e) \big\} \notag \\
&+ \tfrac{12}{5}g_1^2\lamsq(\mhdsq - \mhusq)\mathbf{1}\,.
\end{align}

\subsection{Tadpole Terms}
The general RGE for a SUSY-conserving tadpole term reads
\begin{equation}
\dt L^i = L^p\Gamma_p^i\,,
\end{equation}
and thus for $i=S$ one has
\begin{equation}
\dt \xi_F = \xi_F \Gamma_S^S\,.
\end{equation}
For the soft SUSY-breaking term $\xi_S$, we use the general RGE from \cite{Yam94} because Martin and Vaughn \cite{MV94} do not include the tadpole as part of $\mathcal{L}_{\mathrm{soft}}$.  The relevant RGE reads
\begin{equation}
\dt \xi_S = \frac{1}{16\pi^2} \beta_{\xi_S}^{(1)} + \frac{1}{(16\pi^2)^2} \beta_{\xi_S}^{(2)}\,,
\end{equation}
where the one-loop $\beta$ function is given by
\begin{align}
  \beta_{\xi^{}_S}^{(1)} &= 2(\lamsq + \kapsq)\xi^{}_S + 4(\lambda a_\lambda + \kappa a_\kappa)\xi^{}_F + 2\mu'(2\lambda m_3^2 + \kappa m_S'^2) \notag \\
&+ 4[\lambda\mu (\mhusq + \mhdsq) + \kappa\mu' \mssq] + 4a_\lambda m_3^2 + 2a_\kappa m_S'^2\,.
\end{align}
At two-loop we obtain
\begin{align}
  \beta_{\xi^{}_S}^{(2)} =& -4\lambda^4\big\{ \xi_S + 4(\Alam)\xi_F\big\}
  - 8\kappa^4\big\{ \xi_S + 4(\Akap)\xi_F\big\}
  -6\lamsq\big\{ \xi_S\tr( Y_u Y_u^\dagger)+ 2[(\Alam)\tr( Y_u Y_u^\dagger) + \tr( U_A Y_u^\dagger)] \xi_F \big\} \notag \\
  &-6\lamsq\big\{\xi_S\tr( Y_d Y_d^\dagger) + 2[(\Alam)\tr( Y_d Y_d^\dagger) + \tr( D_A Y_d^\dagger)] \xi_F\big\}  \notag \\
  &-2\lamsq\big\{\xi_S\tr( Y_e Y_e^\dagger) + 2[(\Alam)\tr( Y_e Y_e^\dagger) + \tr( E_A Y_e^\dagger)] \xi_F \big\} \notag \\
  &-8\lamsq\kapsq\big\{ \xi_S + 2[(\Alam) + (\Akap)]\xi_F\big\} \notag \\
  &-12\lambda\big\{ \mtrisq[(\Alam) + \mu']\tr( Y_u Y_u^\dagger) + \mtrisq\tr( U_A Y_u^\dagger)
  + \mu\{\Musq + [(\Alam) + \mu']\tr( U_A Y_u^\dagger) + [\mhdsq + \mhusq]\tr( Y_u Y_u^\dagger)\} \big\} \notag \\
  &-12\lambda\big\{ \mtrisq[(\Alam) + \mu']\tr( Y_d Y_d^\dagger) + \mtrisq\tr( D_A Y_d^\dagger)
  + \mu\{\Mdsq + [(\Alam) +\mu']\tr( D_A Y_d^\dagger) + [\mhdsq + \mhusq]\tr( Y_d Y_d^\dagger)\} \big\}  \notag \\
  &-4\lambda\big\{ \mtrisq[(\Alam)+ \mu']\tr( Y_e Y_e^\dagger) + \mtrisq\tr( E_A Y_e^\dagger)
  + \mu\{\Mesq + [(\Alam) +\mu']\tr( E_A Y_e^\dagger) + [\mhdsq + \mhusq]\tr( Y_e Y_e^\dagger)\} \big\} \notag \\
  &-8\lambda^3\big\{ \mtrisq[2(\Alam) + \mu'] + \mu[\mlamsq + (\Alam)[(\Alam) + \mu'] + \mhdsq + \mhusq]\big\} \notag \\
  &-8\lamsq\kappa\big\{ \msprsq[(\Alam) + (\Akap) + \mu']
  + \mu'[\mlamsq + (\Alam)[(\Akap) + \mu'] + 2\mssq] \big\} \notag \\
  &-8\kappa^3\big\{ \msprsq[2(\Akap) + \mu']
  + \mu'[\mkapsq + (\Akap)[(\Akap) + \mu'] + 2\mssq] \big\} \notag \\
  &+ \tfrac{6}{5}\lambda g_1^2\big\{ 2\mtrisq[(\Alam) + \mu' - M_1]
  + 2\mu[\mhdsq + \mhusq -(\Alam)M_1 - \mu' M_1 + 2M_1^2]
  + \lambda[2\xi_F[(\Alam) - M_1]  + \xi_S] \big\} \notag \\
  &+6\lambda g_2^2\big\{ 2\mtrisq[(\Alam) + \mu' - M_2]
  + 2\mu[\mhdsq + \mhusq - (\Alam)M_2 - \mu' M_2 + 2M_2^2]
  + \lambda[2\xi_F[(\Alam) - M_2] + \xi_S] \big\}\,.
\end{align}

\subsection{Additional Parameters}
Here we list the $\lambda$ and $\kappa$ contributions to the RGEs for the scalar masses $\mtrisq \equiv B\mu$ and $\msprsq \equiv B'\mu'$, and the evolution of the Higgs VEVs $v_{1,2,s}$.  For the former, we get at one-loop
\begin{equation}
\left.\beta_{\mtrisq}^{(1)}\right|_\lambda = 2\lambda(3\lambda\mtrisq + 2\mu a_\lambda) + 2\lambda\kappa\msprsq\,.
\end{equation}
At two-loop we have
\begin{align}
\left.\beta_{\mtrisq}^{(2)}\right|_\lambda &= -2\lambda^4(7\mtrisq + 16\mu a_\lambda/\lambda) -3\lamsq\big\{ 5\mtrisq\tr( Y_u Y_u^\dagger) 
+ 2\mu[3\tr( U_A Y_u^\dagger) + (a_\lambda/\lambda)\tr( Y_u Y_u^\dagger)] \big\} \notag \\
&-3\lamsq\big\{ 5\mtrisq\tr( Y_d Y_d^\dagger) + 2\mu[3\tr( D_A Y_d^\dagger) 
+ (a_\lambda/\lambda)\tr( Y_d Y_d^\dagger) \big\} \notag \\
&-\lamsq\big\{5\mtrisq\tr( Y_e Y_e^\dagger) 
+ 2\mu[3\tr( E_A Y_e^\dagger) + (a_\lambda/\lambda)\tr( Y_e Y_e^\dagger)] \big\} \notag \\
&-4\lamsq\kapsq\big\{ \mtrisq + 2\mu[(a_\lambda/\lambda) + (a_\kappa/\kappa)] \big\} 
-8\lambda^3\kappa\big\{ m_S'^2 + \mu'(a_\lambda/\lambda)\big\} - 8\lambda\kappa^3\big\{m_S'^2 + \mu'(a_\kappa/\kappa)\big\} \notag \\
&+\tfrac{12}{5}g_1^2\lamsq(\mtrisq - \mu M_1) + 12g_2^2\lamsq(\mtrisq -\mu M_2) \,.
\end{align}

For $\msprsq$, the one-loop $\beta$ function reads
\begin{equation}
\beta_{\msprsq}^{(1)} = 4\lambda(\lambda\msprsq + 2\mu' a_\lambda) + 8\kappa(\kappa\msprsq + \mu' a_\kappa) + 8\lambda\kappa\mtrisq \,.
\end{equation}
At two-loop we have
\begin{align}
\beta_{\msprsq}^{(2)} =& -8\lambda^4\big\{\msprsq + 4\mu'(a_\lambda/\lambda)\big\} - 16\kappa^4\big\{ 2\msprsq + 5\mu'(a_\kappa/\kappa) \big\} 
- 16\lamsq\kapsq\big\{ 2\msprsq + \mu'[3(a_\lambda/\lambda) + 2(a_\kappa/\kappa)] \big\} \notag \\
&-12\lamsq\big\{ \msprsq\tr( Y_u Y_u^\dagger) + 2\mu'[(\Alam)\tr( Y_u Y_u^\dagger) + \tr( U_A Y_u^\dagger)] \big\} \notag \\
&-12\lamsq\big\{\msprsq\tr( Y_d Y_d^\dagger) + 2\mu'[(\Alam)\tr( Y_d Y_d^\dagger) + \tr( D_A Y_d^\dagger)] \big\} \notag \\
&-4\lamsq\big\{\msprsq\tr( Y_e Y_e^\dagger) + 2\mu'[(\Alam)\tr( Y_e Y_e^\dagger) + \tr( E_A Y_e^\dagger)] \big\} 
-16\lambda^3\kappa\big\{ \mtrisq + \mu(\Alam)\big \} \notag \\
&-24\lambda\kappa\big\{ \mtrisq\tr( Y_u Y_u^\dagger) + \mu\tr( U_A Y_u^\dagger)\big\} 
-24\lambda\kappa\big\{ \mtrisq\tr( Y_d Y_d^\dagger) + \mu\tr( D_A Y_d^\dagger)\big\} \notag \\
&-8\lambda\kappa\big\{ \mtrisq\tr( Y_e Y_e^\dagger) + \mu\tr( E_A Y_e^\dagger)\big\} 
+\tfrac{24}{5}\lambda\kappa g_1^2\big\{\mtrisq - \mu M_1\big\} + 24\lambda\kappa g_2^2\big\{\mtrisq - \mu M_2 \big\} \notag \\
&+\tfrac{12}{5}\lamsq g_1^2\big\{\msprsq + 2\mu'[(\Alam) - M_1]\big\} 
+12\lamsq g_2^2 \big\{\msprsq + 2\mu'[(\Alam) - M_2] \big\}\,.
\end{align}

At one-loop, the up- and down-type Higgs VEVs $v_{u,d}$ receive additional contributions solely from $\lambda$ \cite{Sper13},
\be
\left.\beta^{(1)}_{v_\alpha}\right|_\lambda = -v_\alpha\lamsq\,, \qquad \alpha=1,2\,,
\ee
while the $\beta$ function for the singlet VEV $s$ is given by
\be 
\beta^{(1)}_s = -2s(\lamsq + \kapsq)\,.
\ee
At two-loop, the $\beta$ functions are given by \cite{Sper13,Sper13-2}
\begin{align}
\beta^{(2)}_{v_1} &= v_1\Bigg\{\gamma^{(2)H_1}_{H_1} - \Big(\tfrac{3}{10}g_1^2 + \tfrac{3}{2}g_2^2\Big)\Big[3\tr(Y_DY_D^\dagger) + \tr(Y_EY_E^\dagger) + \lamsq\Big] + \tfrac{9}{2}g_2^4\Bigg\}\,, \\
\beta^{(2)}_{v_2} &= v_2\Bigg\{\gamma^{(2)H_2}_{H_2} - 
\Big(\tfrac{3}{10}g_1^2 + \tfrac{3}{2}g_2^2\Big)
\Big[3\tr(Y_UY_U^\dagger) + \lamsq\Big] + \tfrac{9}{2}g_2^4\Bigg\}\,, \\
\beta^{(2)}_s &= s\gamma^{(2)S}_S\,.
\end{align}

The one-loop $\beta$ function for $\tan\beta$ is the same in the NMSSM as the MSSM.  At two-loop, one has
\be
\beta^{(2)}_{t_\beta} = \tan\beta\Bigg\{\gamma^{(2)H_2}_{H_2} - 
\gamma^{(2)H_1}_{H_1} + \Big(\tfrac{3}{10}g_1^2 + \tfrac{3}{2}g_2^2\Big)
\frac{\beta^{(1)}_{t_\beta}}{\tan\beta}\Bigg\}\,.
\ee

\end{document}